\documentclass[longauth]{aa}  
\usepackage{array}
\usepackage{graphicx}
\usepackage{txfonts}
\usepackage{natbib,twoopt}
\usepackage{booktabs}
\usepackage{graphicx}
\usepackage{subcaption}
\usepackage[colorlinks=true,linkcolor=blue,citecolor=blue,urlcolor=blue]{hyperref}

\usepackage{tikz}   
\usetikzlibrary{arrows.meta, positioning, shapes.geometric}

\begin{document} 

   \title{The evolutionary history of ultra-compact accreting binaries}

   \subtitle{I. Chemical abundances and formation channel of the eclipsing AM\,CVn system ZTF\,J225237.05$-$051917.4 from \textit{HST} Spectroscopy}

   \author{W.~Yu\inst{1}
          \and
          A.~F.~Pala\inst{2}
          \and 
          T.~Kupfer\inst{1,3}
          \and
          B.~T.~G\"ansicke\inst{4}
          \and
          D.~Koester\inst{5}
          \and
          D.~Belloni\inst{6}
          \and
          T.~L.~S.~Wong\inst{7,8}
          \and
          M.~R.~Schreiber\inst{9}
          \and
          J.~van~Roestel\inst{10, 11}
          \and
          A.~J.~Brown\inst{1}
          \and
          E.~O.~Waagen\inst{12}
          \and
          J.-L.~González-Carballo\inst{13}
          \and
          S.~Bednarz\inst{14}
          \and
          K.~Bernacki\inst{14}
          \and
          D.~De~Martino\inst{15}
          \and
          E.~Fernández~Mañanes\inst{16}
          \and
          R.~González~Farfán\inst{17}
          \and
          M.~J.~Green\inst{18, 19}
          \and
          P.~J.~Groot\inst{20, 21, 22}
          \and
          F.-J~Hambsch\inst{23, 24, 25}
          \and
          C.~Knigge\inst{26}
          \and
          J.-L.~Martin-Velasco\inst{27}
          \and
          M.~Morales-Aimar\inst{28}
          \and
          G.~Myers\inst{29}
          \and
          R.~Naves~Nogues\inst{30}
          \and
          R.~Poggiani\inst{31}
          \and
          A.~Popowicz\inst{14}
          \and
          G.~Ramsay\inst{32}
          \and
          E.~Reina-Lorenz\inst{33}
          \and
          P.~Rodríguez-Gil\inst{34, 35}
          \and
          J.-L.~Salto-González\inst{36}
          \and
          E.~M.~Sion\inst{37}
          \and
          D.~Steeghs\inst{4}
          \and
          P.~Szkody\inst{38}
          \and
          O.~Toloza\inst{9}
          \and
          G.~Tovmassian\inst{39}
          }

   \institute{Hamburg Observatory, University of Hamburg,
              Gojenbergsweg 112, 21029 Hamburg, Germany\\
              \email{weitian.yu@uni-hamburg.de}
              \and
              European Southern Observatory, Karl-Schwarzschild-Strasse 2, 85748 Garching bei München, Germany 
              \and
              Department of Physics and Astronomy, Texas Tech University, 2500 Broadway, Lubbock, TX 79409, USA 
              \and
              Department of Physics, University of Warwick, Coventry, CV4 7AL, UK 
              \and
              Institut für Theoretische Physik und Astrophysik, University of Kiel, 24098 Kiel, Germany 
              \and
              São Paulo State University (UNESP), School of Engineering and Sciences, Guaratinguetá, Brazil 
              \and
              The Observatories of the Carnegie Institution for Science, Pasadena, CA 91101, USA 
              \and
              Department of Physics, University of California, Santa Barbara, CA 93106, USA 
              \and
              Departamento de F\'isica, Universidad T\'ecnica Federico Santa Mar\'ia, Av. Espa\~na 1680, Valpara\'iso, Chile 
              \and
              Institute of Science and Technology Austria, Am Campus 1, 3400, Klosterneuburg, Austria 
              \and
              Anton Pannekoek Institute for Astronomy, University of Amsterdam, 1090 GE Amsterdam, The Netherlands 
              \and
              American Association of Variable Star Observers (AAVSO), 185 Alewife Brook Pkwy, Suite 410, Cambridge, MA 02138, USA 
              \and
              Observadores de Supernovas (ObSN), Observatorio Cerro del Viento, MPC I84, Pl. Fernández Pirfano, 3-5A, Badajoz, 06010, Spain 
              \and
              Silesian University of Technology, Akademicka 16, Gliwice, Poland 
              \and
              INAF - Osservatorio Astronomico di Capodimonte, Salita Moiariello 16, 80131, Naples, Italy 
              \and
              Observadores de Supernovas (ObSN), Observatorio Estelia, MPC Y90, C/ Ladines, 12, Ladines, Asturias, 33993, Spain 
              \and
              Observadores de Supernovas (ObSN), Observatorio Uraniborg  C/ Antequera, 8, 41400 Écija, Sevilla, Spain 
              \and
              Homer L. Dodge Department of Physics and Astronomy, University of Oklahoma, 440 W. Brooks Street, Norman, OK 73019, USA 
              \and
              JILA, University of Colorado and National Institute of Standards and Technology, 440 UCB, Boulder, CO 80309-0440, USA 
              \and
              Department of Astrophysics/IMAPP, Radboud University, PO Box 9010, 6500 GL Nijmegen, The Netherlands 
              \and
              Department of Astronomy, University of Cape Town, Private Bag X3, Rondebosch 7701, South Africa 
              \and
              South African Astronomical Observatory, PO Box 9, Observatory 7935, South Africa 
              \and
              Vereniging Voor Sterrenkunde (VVS), Zeeweg 96, 8200 Brugge, Belgium 
              \and
              Groupe Européen d’Observations Stellaires (GEOS), 23 Parc de Levesville, 28300 Bailleau l'Eveque, France 
              \and
              Bundesdeutsche Arbeitsgemeinschaft für Veränderliche Sterne (BAV), Munsterdamm 90, 12169 Berlin, Germany 
              \and
              School of Physics and Astronomy, University of Southampton, Highfield, Southampton SO17 1BJ, UK 
              \and
              Observadores de Supernovas (ObSN), Observatorio Carpe Noctem, MPC I72, Paseo de la Maliciosa 11, Collado Mediano, Madrid, 28027, Spain 
              \and
              Observadores de Supernovas (ObSN), Observatorio de Sencelles, MPC K14, Camí de Sonfred 1, Sencelles, Islas Baleares, 07140, Spain 
              \and
              American Association of Variable Star Observers (AAVSO), 5 Inverness Way, Hillsborough, CA 94010 USA 
              \and
              Observadores de Supernovas (ObSN), Observatorio Montcabrer, MPC 213, C/ Jaume Balmes 24, Cabrils, Barcelona, 08348, Spain 
              \and
              Dipartimento di Fisica, Universit\`a di Pisa, 56127 Pisa, Italy 
              \and
              Armagh Observatory \& Planetarium, College Hill, Armagh BT61 9DG, UK 
              \and
              Observadores de Supernovas (ObSN), Observatorio de Masquefa, MPC 232, Av. Can Marcet 41, Masquefa, Barcelona, 08783, Spain 
              \and
              Instituto de Astrofísica de Canarias, E-38205 La Laguna, Tenerife, Spain 
              \and
              Departamento de Astrofísica, Universidad de La Laguna, E-38206 La Laguna, Tenerife, Spain 
              \and
              Observadores de Supernovas (ObSN), Cal Maciarol mòdul 8 Observatory, MPC A02, Masia Cal Maciarol, Camí de l'Observatori s/n, Àger, Lleida, 25691, Spain 
              \and
              Department of Astrophysics and Planetary Science, Villanova University, Villanova, PA 19085, USA 
              \and
              Department of Astronomy, University of Washington, Seattle, WA 98195, USA 
              \and
              Instituto de Astronom\'ia, Universidad Nacional Aut\'onoma de M\'exico, Ciudad Universitaria 04510, CDMX, M\'exico. 
             }

   \date{Received October 06, 2025; accepted November 30, 2025}
 
\abstract
    {AM\,Canum Venaticorum (AM\,CVn) stars are ultra-compact binary systems composed of a white dwarf primary accreting from a hydrogen-deficient donor. They play a crucial role in astrophysics as potential progenitors of Type~Ia supernovae and as laboratories for gravitational wave studies. However, their formation and evolutionary history remain unsolved. Three formation channels have been discussed in the literature: the white dwarf, the He-star, and the cataclysmic variable channel.}
    {The chemical composition of the accretor atmosphere reflects the material transferred from the donor. In this work, we aim to provide the first accurate measurements of the fundamental parameters of the accreting white dwarf in ZTF\,J225237.05$-$051917.4, including the abundances of key elements such as carbon, nitrogen, and silicon, by analysing ultraviolet spectra obtained with the \textit{Hubble Space Telescope} (\textit{HST}). These measurements provide new insight into the evolutionary history of the system and, together with existing optical observations, establish it as a benchmark to develop our pipeline, paving the way for its application to a larger sample of AM\,CVns.}
    {We determine the binary parameters through photometric analysis and constrain the atmospheric parameters of the white dwarf accretor, including effective temperature, surface gravity, and chemical abundances, by fitting the \textit{HST} ultraviolet spectrum with synthetic spectral models. We then infer the system’s formation channel by comparing the results with theoretical evolutionary models.}
    {We measure the accretor’s effective temperature $T_{\mathrm{eff}} = 23\,300 \pm 600$\,K and surface gravity $\log g = 8.4 \pm 0.3$, which imply an accretor mass of $M_{\mathrm{WD}} = 0.86 \pm 0.16\,M_{\odot}$. We find a high nitrogen-to-carbon abundance ratio by mass of $>153$.}
    {The accretor is significantly hotter than previous estimates based on simplified blackbody fits to the spectral energy distribution, underscoring the importance of detailed spectral modelling for determining accurate system parameters. Our results show that ultraviolet spectroscopy is well-suited to constraining the formation channels of AM\,CVns. Among the three proposed formation channels, the He-star channel can be excluded given the high nitrogen-to-carbon ratio. Our results are consistent with both the white dwarf and cataclysmic variable channels.}


   \maketitle

\section{Introduction} \label{Introduction}

    AM\,Canum\,Venaticorum (AM\,CVn) stars are ultra-compact accreting binary systems composed of a white dwarf (WD) primary and a hydrogen-deficient donor. Depending on their evolutionary history, donor stars may be WDs or semi-degenerate He-stars. These systems are crucial in several areas of astrophysics: (1) they serve as probes of stellar evolution, representing the final stages of close binary evolution; (2) they are potential progenitors of Type Ia supernovae (SNe\,Ia); and (3) they act as laboratories for gravitational wave (GW) astrophysics. The number of known AM\,CVns has increased significantly, from a few dozen in the early 2000s to more than one hundred, mainly due to large surveys \citep{Roelofs_2007, Solheim_2010, Ramsay_2018, Pichardo_Marcano_2021, Green_2025}. With ongoing and upcoming surveys, including the BlackGEM project \citep{Groot_2024}, the Gravitational-wave Optical Transient Observer \citep[GOTO;][]{Steeghs_2022}, the Sloan Digital Sky Survey - V \citep[SDSS;][]{Kollmeier_2025}, and the Legacy Survey of Space and Time \citep[LSST;][]{Ivezic_2019}, many more are expected to be discovered. With the growing sample, it is timely to develop a more detailed understanding of these systems.
    
    An AM\,CVn system descends from binary main-sequence stars, in which the more massive component evolves first off the main sequence and enters the giant phase. This leads to a binary interaction event, which may proceed through either stable Roche-lobe overflow or a common envelope phase \citep{Brown_2016, Li_2023}. This interaction episode leaves behind a more compact binary with a WD primary. Three widely accepted evolutionary channels may follow: 
    (i) the WD channel, where the secondary evolves off the main sequence and the system undergoes a second interaction, leading to a double WD binary. GW radiation then shrinks the orbit to periods as short as five minutes, after which stable mass transfer begins \citep{Paczynski_1967, Webbink_1984, Nelemans_2001, Deloye_2007, Wong_2021, Chen_2022}; 
    (ii) the He-star channel, which also involves a second interaction episode but leaves behind a semi-degenerate, helium-rich donor. In this case, the initially non-degenerate donor, being less compact, causes mass transfer at longer orbital periods, typically around 15 minutes \citep{Savonije_1986, Iben_1987, Yungelson_2008, Bauer_2021}; and 
    (iii) the cataclysmic variable (CV) channel, which differs from the other two as it originates in a CV system and involves only one interaction phase. The helium core of the donor is exposed once its outer layers are stripped through stable mass transfer, during which the accreted material transitions from hydrogen-rich to helium-rich \citep{Tutukov_1985, Podsiadlowski_2003, Goliasch_2015, Liu_2021, Belloni_2023, Sarkar_2023}.

    A possible final fate of an AM\,CVn is the explosive event of a SN\,Ia, one of the most important classes of astronomical transients. In the so-called "double-detonation" scenario, a surface helium detonation occurs on a WD after it has accreted sufficient material from a helium-rich donor \citep{Nomoto_1982, Bildsten_2007, Fink_2010, Shen_2018a, Shen_2018b, Wang_2018, Wong_2023, Rajamuthukumar_2024}. When this helium layer reaches a critical mass, it ignites and can trigger a secondary detonation in the carbon--oxygen (CO) core of the WD, resulting in a thermonuclear explosion. This scenario is particularly relevant for sub-Chandrasekhar mass WDs, as it does not require the WD to exceed the Chandrasekhar limit. A double-shell morphology of a supernova remnant was recently reported by \citet{Das_2025}, providing observational evidence of the existence of this explosion mechanism in nature.

    Owing to their compact orbits and relatively small distances to Earth, AM\,CVn systems are among the strongest sources of GWs in the millihertz band, the operating range of the forthcoming \textit{Laser Interferometer Space Antenna} (LISA) and \textit{TianQin} missions \citep{Amaro-Seoane_2023, Luo_2016}. Many AM\,CVn binaries will be detectable within weeks to months after the launch of \textit{LISA}, serving as verification sources for instrument calibration and early science \citep{Nelemans_2004, Kremer_2017, Kupfer_2018}. \citet{Kupfer_2024} identified 40 Galactic binaries likely to be detected by \textit{LISA}, including 18 verification sources.\footnote{\citet{Kupfer_2024} defines a detectable binary as one that can be identified by \textit{LISA} within 48 months of operation, and a verification source as one detected within the first three months. Detectability is assessed from the shape and confidence of the recovered posterior distributions of the binary parameters.} Among these, 16 AM\,CVn systems are expected to be detected after 48 months, and two may be resolved within just two months. With the anticipated launch of \textit{LISA} in the 2030s, detailed electromagnetic characterisation of these systems is essential. Such efforts will improve the interpretation of GW sources, enable multi-messenger studies, and enhance the calibration of GW observatories. Systems that are not individually resolved will still contribute to the Galactic confusion foreground, produced by compact binaries in the Milky Way, which can limit sensitivity to weak signals \citep{Breivik_2020, Korol_2022, Littenberg_2025}.
    
    A critical question in the study of AM\,CVn systems is the relative contribution of their formation channels. This constrains binary population synthesis predictions for their space densities and period distributions. One possible way to distinguish between the channels is through the chemical composition of the donor \citep{Nelemans_2010}. While in a few systems, there are indications that the donor may be detectable in the infrared \citep{Green_2020, Rivera_Sandoval_2021}, donors are more commonly intrinsically faint and challenging to observe directly. Their composition can therefore be inferred only from the surface abundances of the accreting WD, which trace the material stripped from the donor. This indirect approach enables reconstruction of the evolutionary history of the system. In particular, the N/O and N/C ratios vary strongly due to different levels of CNO and He burning. The key diagnostic is the N/C ratio, which differs by more than an order of magnitude: systems with a He-star donor show N/C $\lesssim 10$, whereas both the WD–WD and CV channels exhibit N/C $\gtrsim 100$, highlighting the importance of carbon detection. Models for the CV channel have traditionally struggled to reproduce the observed systems, as they predict a modest amount of residual hydrogen \citep{Schenker_2002, Podsiadlowski_2003}. Recent studies, however, suggest that modifications such as enhanced magnetic braking may resolve these difficulties, potentially allowing donors to retain hydrogen at levels below current detection thresholds \citep{Belloni_2023, Sarkar_2023}.

    High-quality ultraviolet (UV) spectroscopy is essential for detailed stellar atmosphere analysis of compact accreting binaries. Optical observations alone are insufficient, as the accretion disc dominates the system flux and outshines the WD in the optical and (near-)infrared band. Moreover, absorption features of key elements such as silicon, nitrogen, carbon, and oxygen are primarily detectable in the UV. Space-based spectroscopy is therefore the only means of accurately determining the system parameters such as effective temperature ($T_{\mathrm{eff}}$), surface gravity ($\log g$), and elemental abundances. The only currently operational telescope capable of such observations is the \textit{Hubble Space Telescope} (\textit{HST}). In response to this need, a large \textit{HST} Treasury observing campaign (PI: A.~F.~Pala) was carried out, targeting 31 accreting WD binaries, including 12 AM\,CVns.

    Among the AM\,CVns observed with \textit{HST}, we focus on the eclipsing system ZTF\,J225237.05$-$051917.4 (hereafter ZTF\,J2252$-$05) with orbital period 37.4 min as a benchmark target. This system was discovered by \citet{van_Roestel_2022}, who also carried out a photometric and spectroscopic follow-up. Observations from the Zwicky Transient Facility (ZTF; \citealt{Bellm_2019}), together with high-speed photometry with Chimera on the Hale Telescope enabled precise measurements of the orbital period ($P_{\mathrm{orb}}$) and mass ratio ($q$). In addition, optical spectroscopy obtained with the Low Resolution Imaging Spectrometer (LRIS) on the Keck~I telescope revealed a helium-dominated atmosphere with no detectable hydrogen features. Compared to other AM\,CVns, ZTF\,J2252$-$05 is characterised by one of the most comprehensive observational data sets. It is therefore well-suited to serve as a foundation for developing a pipeline to constrain the physical parameters of such systems spectroscopically.
    
    This paper is structured as follows. Section~\ref{Observation} presents the observational data, including ground-based photometry in support of the \textit{HST} observations, and describes the structure and processing of the time-resolved ultraviolet spectroscopy. Section~\ref{Method} details the methods used for light curve and spectral modelling. The results of the spectral analysis are given in Section~\ref{Result}, and the evolutionary history is discussed in Section~\ref{Discussion}. Section~\ref{Conclusion} summarises our findings and outlines the steps required to build the first statistically significant spectroscopic sample of AM\,CVn stars to study their evolution.


\section{Observations} \label{Observation}

    ZTF\,J2252$-$05 was observed using the Cosmic Origins Spectrograph (COS) onboard \textit{HST} on 2022~November~17 with a total exposure time of 11\,564\,s over five spacecraft orbits. The G140L grating was used with a central wavelength of 800\,\AA, covering 900--2150\,\AA\ at $R\simeq3000$. It was observed as part of a large \textit{HST} Treasury programme (Cycle~29; ID: 16659 and 17401; PI: A.~F.~Pala), which includes 31 accreting white dwarfs, among them 12 AM\,CVns (11 of which have been observed at the time of writing). The observational information of the AM\,CVn targets is summarised in Table~\ref{tab:HST_program}. Orbital periods $P_{\mathrm{orb}}$ (excluding ZTF\,J2252$-$05) are adopted from \citet{Green_2025}. A comprehensive discussion of the AM\,CVn sample, together with the broader set of accreting binaries in the programme, will be presented in future work.
    \begin{table*}
        \caption{AM\,CVn targets observed in \textit{HST} programme 16659 and 17401.}
        \label{tab:HST_program}
        \centering
        \begin{tabular}{l c c c l l}
            \toprule
            Target & RA & DEC & $P_{\mathrm{orb}}$ & Observation date & Exposure time \\
                   &    &     & [min]              & [YYYY-MM-DD]     & [s] \\
            \midrule
            ES\,Cet                     & $02^{\mathrm{h}} 00^{\mathrm{m}} 52.24^{\mathrm{s}}$ 
                                        & $-09^{\circ} 24' 31.64''$        
                                        & 10.3 & 2024-10-24 & 4068 \\

            KIC\,4547333                & $19^{\mathrm{h}} 08^{\mathrm{m}} 17.08^{\mathrm{s}}$ 
                                        & $+39^{\circ} 40' 36.45''$
                                        & 18.2 & 2022-06-03 & 6925 \\
            
            HP\,Lib                     & $15^{\mathrm{h}} 35^{\mathrm{m}} 53.07^{\mathrm{s}}$ 
                                        & $-14^{\circ} 13' 12.21''$        
                                        & 18.4 & 2023-06-23, 2025-03-20 & 2505, 4546 \\
                                        
            YZ\,LMi                     & $09^{\mathrm{h}} 26^{\mathrm{m}} 38.72^{\mathrm{s}}$ 
                                        & $+36^{\circ} 24' 02.47''$
                                        & 28.3 & 2021-12-30 & 9210 \\
                                        
            GALEX\,J113315.3$-$371019   & $11^{\mathrm{h}} 33^{\mathrm{m}} 16.36^{\mathrm{s}}$ 
                                        & $-37^{\circ} 10' 19.95''$        
                                        & 31.2 & 2024-05-17 & 6473 \\

            ZTF\,J040749.30$-$000716.6  & $04^{\mathrm{h}} 07^{\mathrm{m}} 49.30^{\mathrm{s}}$ 
                                        & $-00^{\circ} 07' 16.67''$        
                                        & 35.0 & 2023-02-10, 2024-01-11 & 6600, 6358 \\

            ZTF\,J225237.05$-$051917.4  & $22^{\mathrm{h}} 52^{\mathrm{m}} 37.05^{\mathrm{s}}$ 
                                        & $-05^{\circ} 19' 16.97''$        
                                        & 37.4 & 2022-11-17 & 11564 \\          

            ASASSN-14mv                 & $07^{\mathrm{h}} 13^{\mathrm{m}} 27.28^{\mathrm{s}}$ 
                                        & $+20^{\circ} 55' 53.28''$
                                        & 40.9 & 2022-03-25 & 6406 \\

            SDSS\,J080449.49$+$161624.8 & $08^{\mathrm{h}} 04^{\mathrm{m}} 49.49^{\mathrm{s}}$ 
                                        & $+16^{\circ} 16' 24.87''$        
                                        & 44.5 & 2023-02-16, 2024-02-16 & 2358, 4210 \\                                        
                                        
            GP\,Com                     & $13^{\mathrm{h}} 05^{\mathrm{m}} 42.40^{\mathrm{s}}$ 
                                        & $+18^{\circ} 01' 03.76''$        
                                        & 46.5 & 2022-04-29, 2023-05-25 & 2368, 2368 \\
                                        
            ASASSN-14cn                 & $16^{\mathrm{h}} 11^{\mathrm{m}} 33.97^{\mathrm{s}}$ 
                                        & $+63^{\circ} 08' 31.88''$
                                        & 49.7 & 2022-09-01 & 14206 \\
            \bottomrule
        \end{tabular}
    \end{table*}

    \subsection{Ultraviolet observations} \label{sub.cos_observation}
             
\begin{figure*}
            \centering
            \includegraphics[width=\textwidth]{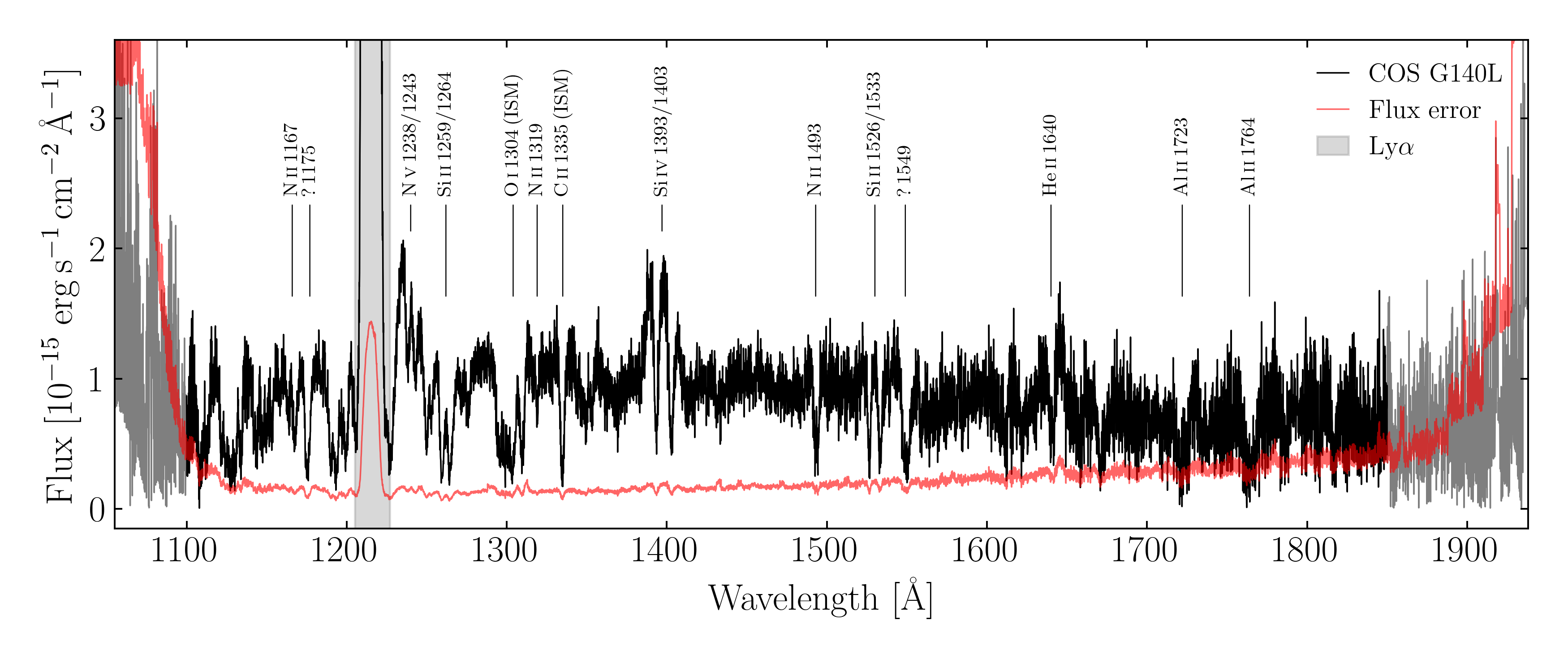}
            \caption{Average \textit{HST}/COS spectrum of ZTF\,J2252$-$05 covering 900--2050\,\AA. Masked regions include areas of low sensitivity ($<1100$ and $>1850$\,\AA; grey) and geocoronal Ly$\alpha$ at 1215\,\AA\ (shaded). Night-only data were used to reduce airglow contamination. The error of the averaged flux is shown in red.}
            \label{fig.ztfj2252_spec}
        \end{figure*}

        The COS-averaged ultraviolet spectrum of ZTF\,J2252$-$05 is shown in Fig.~\ref{fig.ztfj2252_spec}. The usable wavelength range extends from 1100 to 1800\,\AA, where the noise level is sufficiently low (signal-to-noise ratio, S/N $\gtrsim 4$ on the pseudo-continuum). The Geocoronal Ly$\alpha$ contaminated region (1196--1225\,\AA) is masked. To recover the photospheric \ion{O}{I} absorption line at 1304\,\AA, we used only night-time \textit{HST} data, when the Sun was below the Earth limb and geocoronal airglow was minimal. The average spectrum is then obtained by patching the clean O\,\textsc{i} region into the full spectrum, to preserve high S/N elsewhere.

\begin{figure*}
            \centering
            \includegraphics[width=\textwidth]{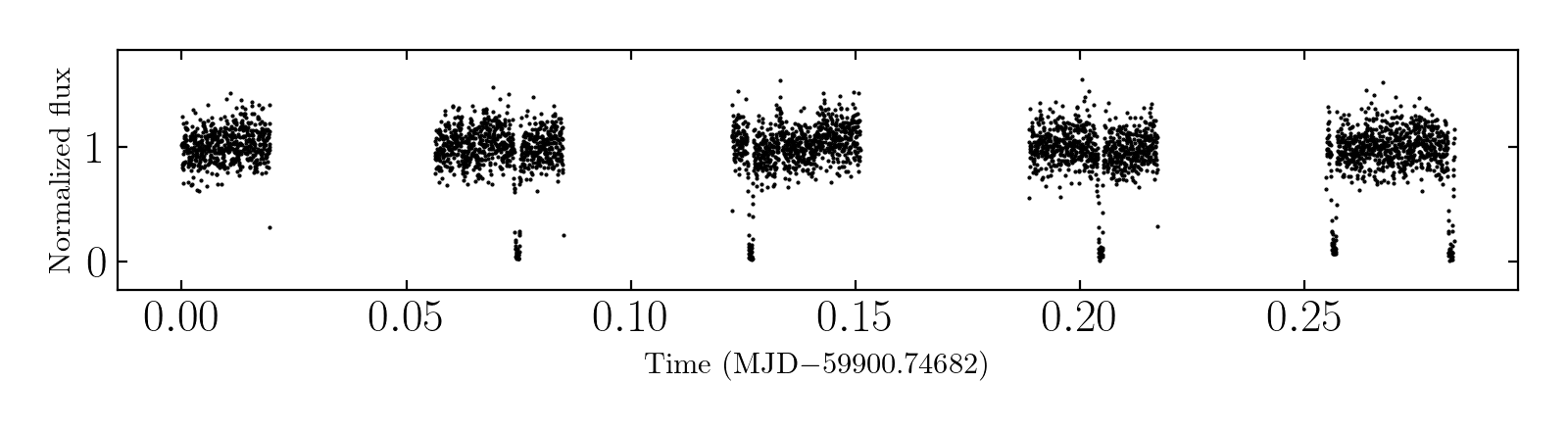}
            \caption{Normalised UV light curve of ZTF\,J2252$-$05, extracted from time-resolved COS observations. The total exposure time was 11\,563\,s, covering five eclipses. The gaps correspond to intervals when the target was behind the Earth and not visible to \textit{HST}.}
            \label{fig.lc_cos}
        \end{figure*}

        The data were obtained in \texttt{TIME-TAG} mode, which also allows us to extract the light curve, following a method similar to that described in \citet{Pala_2022}. By plotting photon positions in the cross-dispersion direction (YFULL) against their corresponding wavelengths, we defined a rectangular region centred on the target spectrum. Two parallel background regions, one on each side of the target spectrum, are also selected for calibration. Photon events were binned into 3\,s intervals. The source signal was extracted by counting events within the rectangular aperture corresponding to the target, and corrected for background counts measured in adjacent regions to produce the light curve (Fig.~\ref{fig.lc_cos}).
        
    \subsection{Brightness monitoring}
    
\begin{figure}
            \centering
            \includegraphics[width=\columnwidth]{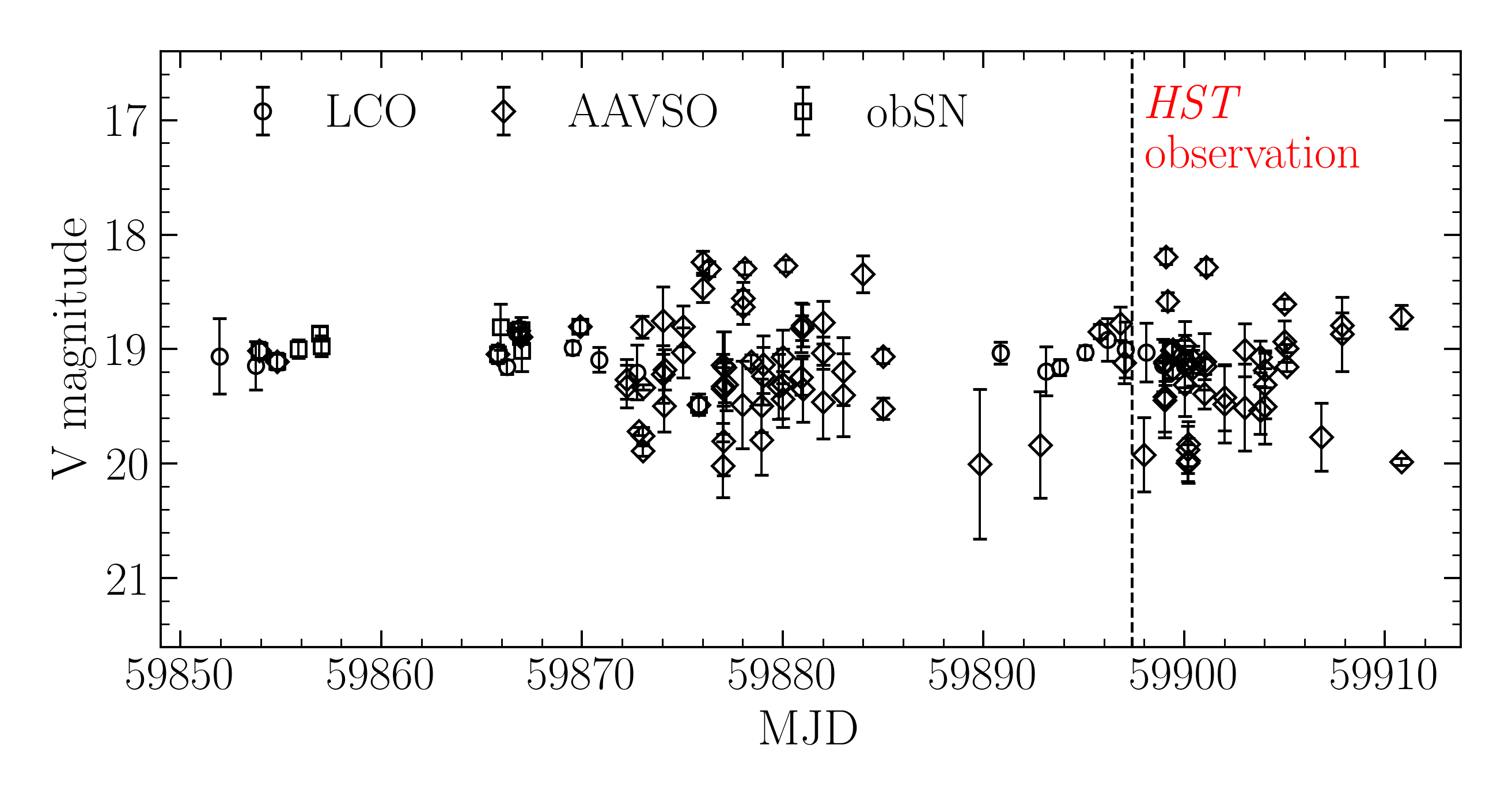}
            \caption{Photometric monitoring of ZTF\,J2252$-$05, including observations from LCO, AAVSO and ObSN. The vertical dashed line highlights the time of the \textit{HST} observations (2022 November 17). Fluctuations can occur during monitoring, as a single exposure may coincide with, or partially cover, an eclipse ($\simeq1.5$\,min) of the WD.}
            \label{fig.LCO_monitoring}
        \end{figure}

        Due to the intrinsic variability of accreting binary systems, observations of accreting WDs are only safe during quiescence. Observing the target during a disc outburst, when the system brightness increases by a factor of 2–8, could exceed the COS safety thresholds and pose a risk to the detectors. In addition, during an outburst, the disc dominates the emission, which prevents the detection of the WD. Following the procedure described in previous works \citep{Pala_2017, Tovmassian_2025}, we therefore conducted a photometric monitoring campaign in the weeks preceding the scheduled \textit{HST} observations to ensure both safety and scientific return. Photometry was obtained using the robotic ground-based telescopes of the Las Cumbres Observatory (LCO; \citealt{Brown_2013}), complemented by coordinated observations from citizen scientists affiliated with the American Association of Variable Star Observers (AAVSO) and the Observadores de Supernovas (ObSN) group. Figure~\ref{fig.LCO_monitoring} shows the brightness of ZTF\,J2252$-$05 in the weeks around the \textit{HST} visit, remaining stable at $V \approx 19$\,mag and indicating that the system was in quiescence.

\section{Methods} \label{Method}

    \subsection{Light curve analysis} \label{sub.lc_analysis}

\begin{figure}
            \centering
            \includegraphics[width=\columnwidth]{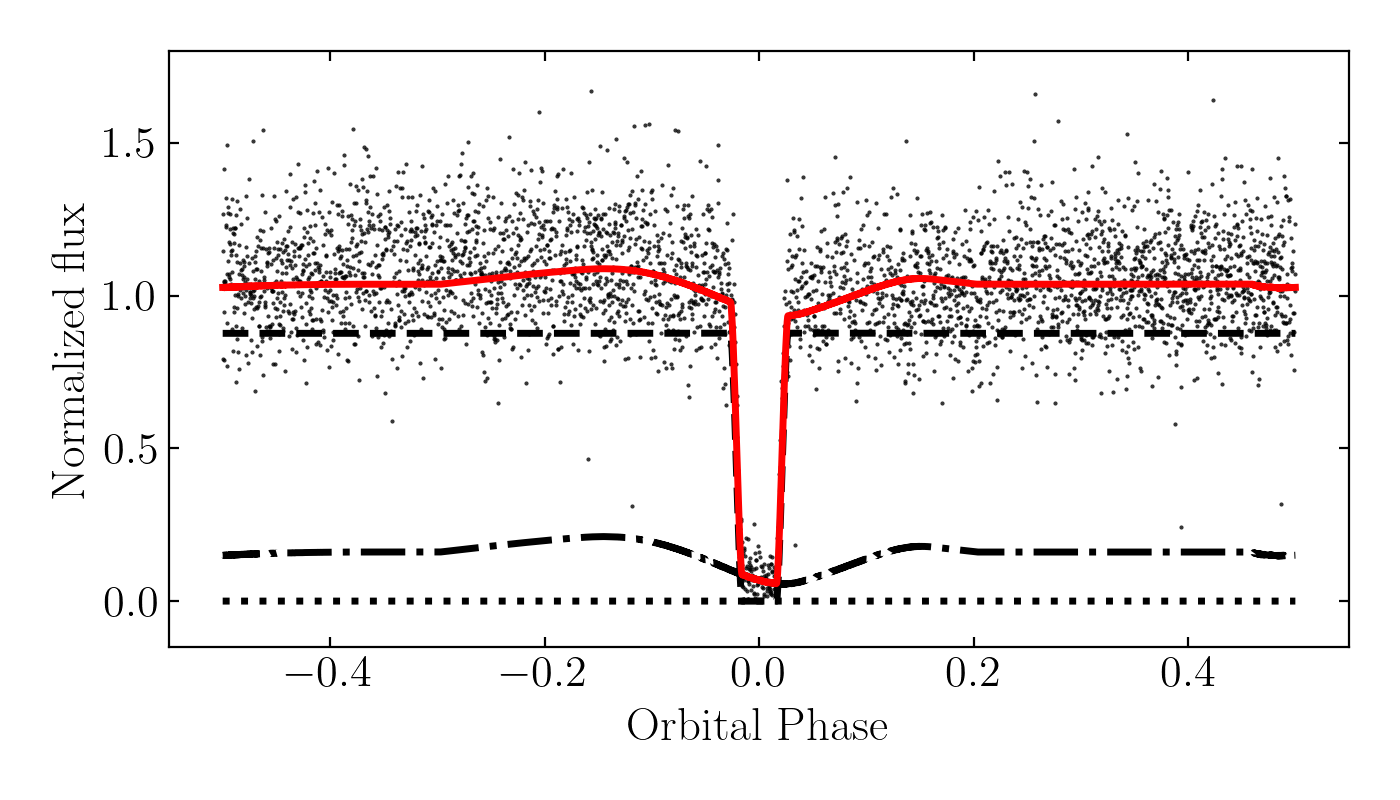}
            \caption{Fitted UV light curve of ZTF\,J2252$-$05. The black points show the light curve binned into 3\,s intervals. The best-fit model (red) includes contributions from the accreting WD (dashed line), the bright spot (dash-dotted line), and the accretion disc (dotted line). The residuals are shown below the light curve.}
            \label{fig.lc_fit}
        \end{figure}

        The light-curve analysis of ZTF\,J2252$-$05 provides a baseline for understanding the system and serves as a reference for the subsequent spectral analysis by constraining the binary parameters. Geocoronal Ly$\alpha$ and \ion{O}{I}, together with strong disc emission features (\ion{N}{V} and \ion{Si}{IV}) that do not originate from the WD, were masked. We model the eclipsing UV light curve with the \textsc{lcurve} software package \citep{Copperwheat_2010}. The best-fitting model is shown in Fig.~\ref{fig.lc_fit}, and the key corresponding parameters are listed in Table~\ref{tab:system_parameters}. We adopt the binary parameters, such as $P_{\mathrm{orb}}$ and $q$, from \citet{van_Roestel_2022}, who derived them from high-time-resolution optical photometry. The donor radius, $R_{\mathrm{donor}}$, was determined from $q$ under the assumption of Roche-lobe filling \citep{Eggleton_1983}. We modelled the individual components of the system, namely the white dwarf, the accretion disc, and the bright spot (the region where the accreting material impacts the outer edge of the disc), following the approach of \citet{Green_2018}. The free parameters of the model are the inclination, WD radius, component temperatures (WD, donor, disc, and bright spot), and the reference epoch $T_{0}$, defined as the mid-eclipse time.
        
        This allows us to estimate the relative contributions of the components to the total flux. The model indicates that the accreting WD contributes $84\,\%$ of the UV flux, the bright spot $16\,\%$, and the disc less than $0.1\,\%$. The disc contribution is therefore negligible and excluded from the subsequent spectral analysis. In the optical, the bright spot becomes increasingly relevant towards shorter wavelengths, contributing $3\,\%$, $5\,\%$, and $7\,\%$ in the $i$, $r$, and $g$ bands, respectively \citep{van_Roestel_2022}. This trend reflects its high temperature ($>14\,000$~K), in contrast to the cooler disc ($\sim 5000$~K), which dominates the optical flux but has no significant impact in the UV.

    \subsection{Spectral analysis} \label{spectral_analysis}
    
        \subsubsection{Interstellar medium contamination} \label{sub.ISM}
    
            Extinction and absorption features from the interstellar medium (ISM) must be accounted for in the analysis of stellar atmospheres. ZTF\,J2252$-$05 is located at Galactic coordinates $(l, b) = (76.96^\circ, -46.67^\circ)$ with a distance of $d = 536^{+82}_{-93}$\,pc. The distance was reported by \citet{van_Roestel_2022}, based on the Gaia~DR3 parallax \citep{Brown_2021} and a prior constructed from the Galactic white dwarf distribution \citep{Kupfer_2018}. The extinction is $A_V = 0.149 \pm 0.007$ from the 3D Galactic dust maps (G-Tomo; \citealt{Lallement_2022, Vergely_2022}), which for $R_V = 3.1$ returns $E(B-V) = 0.048 \pm 0.002$. We then redden our model accordingly following the wavelength-dependent extinction law from \citet{Cardelli_1989}.
            
\begin{figure*}
                \centering
                \includegraphics[width=0.9\textwidth]{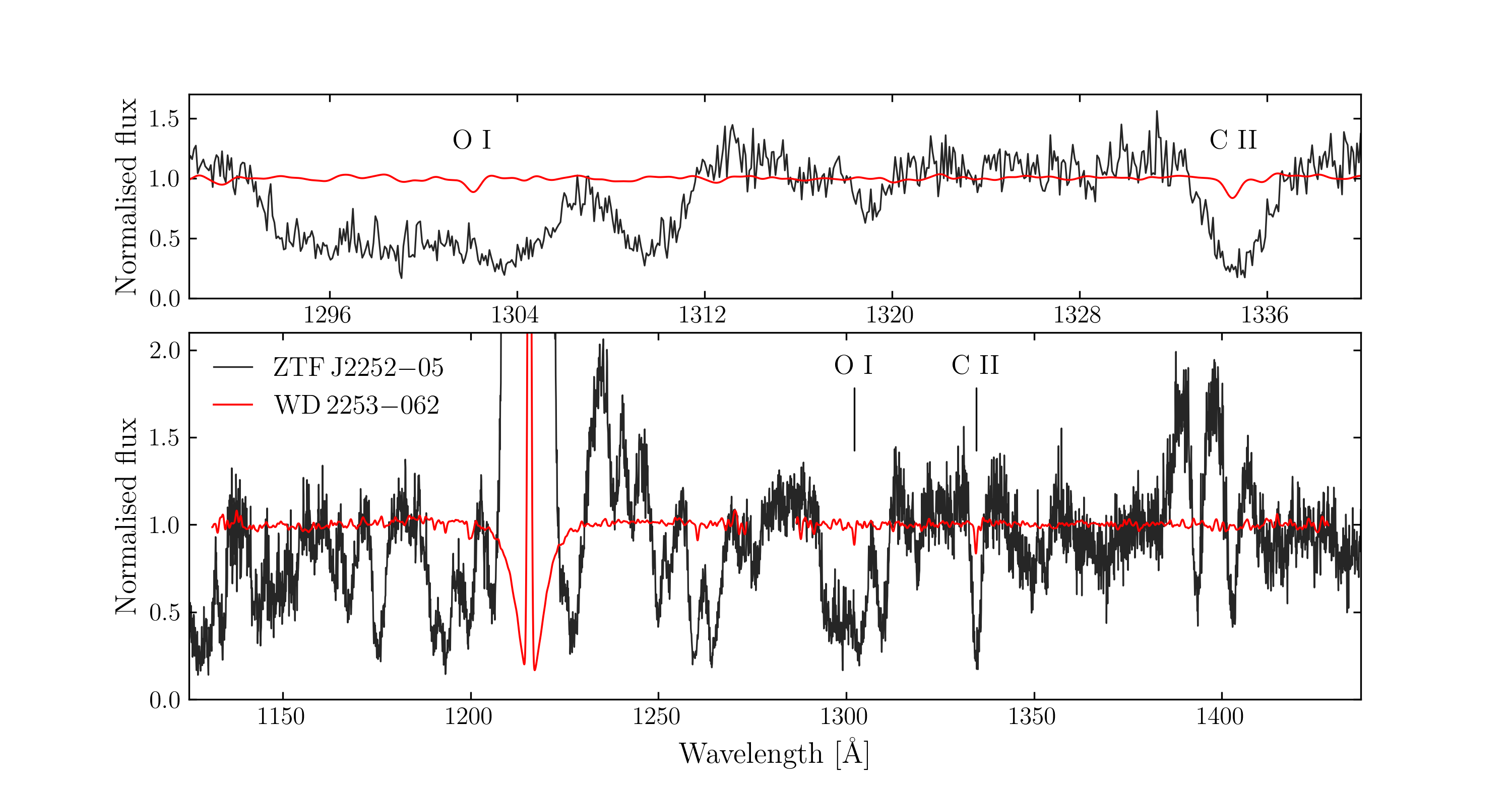}
                \caption{Spectral comparison between ZTF\,J2252$-$05 (black) and WD\,2253$-$062 (red). The lower panel shows the spectrum of WD\,2253$-$062 obtained with the G130M grating, degraded to match the resolution of the ZTF\,J2252$-$05 observation with the G140L grating. The upper panel presents a zoom-in on the potential ISM-contaminated regions around \ion{O}{I}\,$1304$ and \ion{C}{II}\,$1335$\,\AA. Both spectra were continuum normalised.}
                \label{fig.wd2253}
            \end{figure*}

            To assess possible contamination by interstellar metal lines, we compared the spectrum of ZTF\,J2252$-$05 with that of a target WD\,2253$-$062 (RA: 22:55:47.52, Dec: $-06{:}00{:}50.67$; $(l, b) = (65.21^\circ, -55.34^\circ)$; $d = 67.8$\,pc). Being located only $63\arcmin$ away from ZTF\,J2252$-$05, this is the closest WD on the sky with a pristine hydrogen atmosphere and archival \textit{HST} observations. It was observed with COS/G130M for an exposure of 2000\,s (Cycle~25, ID: 15073; PI: B.~G\"ansicke). As shown in Fig.~\ref{fig.wd2253}, the spectra of ZTF\,J2252$-$05 and WD\,2253$-$062 both exhibit absorption lines from oxygen and carbon. The prominent \ion{C}{II} 1335\,\AA\ and \ion{O}{I} 1304\,\AA\ lines, which are also present in WD\,2253$-$062, indicate an interstellar origin. Therefore, these lines are excluded from the following spectral analysis.
            
            Given the importance of carbon and oxygen as tracers of nuclear burning, we examine the alternative ionisation stages. For oxygen, reliable diagnostic lines are not available within the observed wavelength: the \ion{O}{I} $\lambda 1026$\,\AA\ line lies in the noisy far-UV region, and the broad geocoronal Ly$\alpha$ emission saturates the potential ionised $\lambda 1217$\,\AA\ oxygen feature. As a result, we cannot place a meaningful constraint on the oxygen abundance. For carbon, we can estimate an upper limit of its abundance from the theoretical C\,\textsc{i} feature near 1280\,\AA\ (Sect.~\ref{sub.carbon_hydrogen_constrain}). However, even though ZTF\,J2252$-$05 is angularly close to WD\,2253$-$062, it is eight times farther away. Additional ISM contamination cannot be ruled out, and the constraints on carbon from \ion{C}{I} are consequently only upper limits.

        \subsubsection{Spectral fitting} \label{sub.spectral_fit}
    
\begin{figure}
                \centering
                \includegraphics[width=\columnwidth]{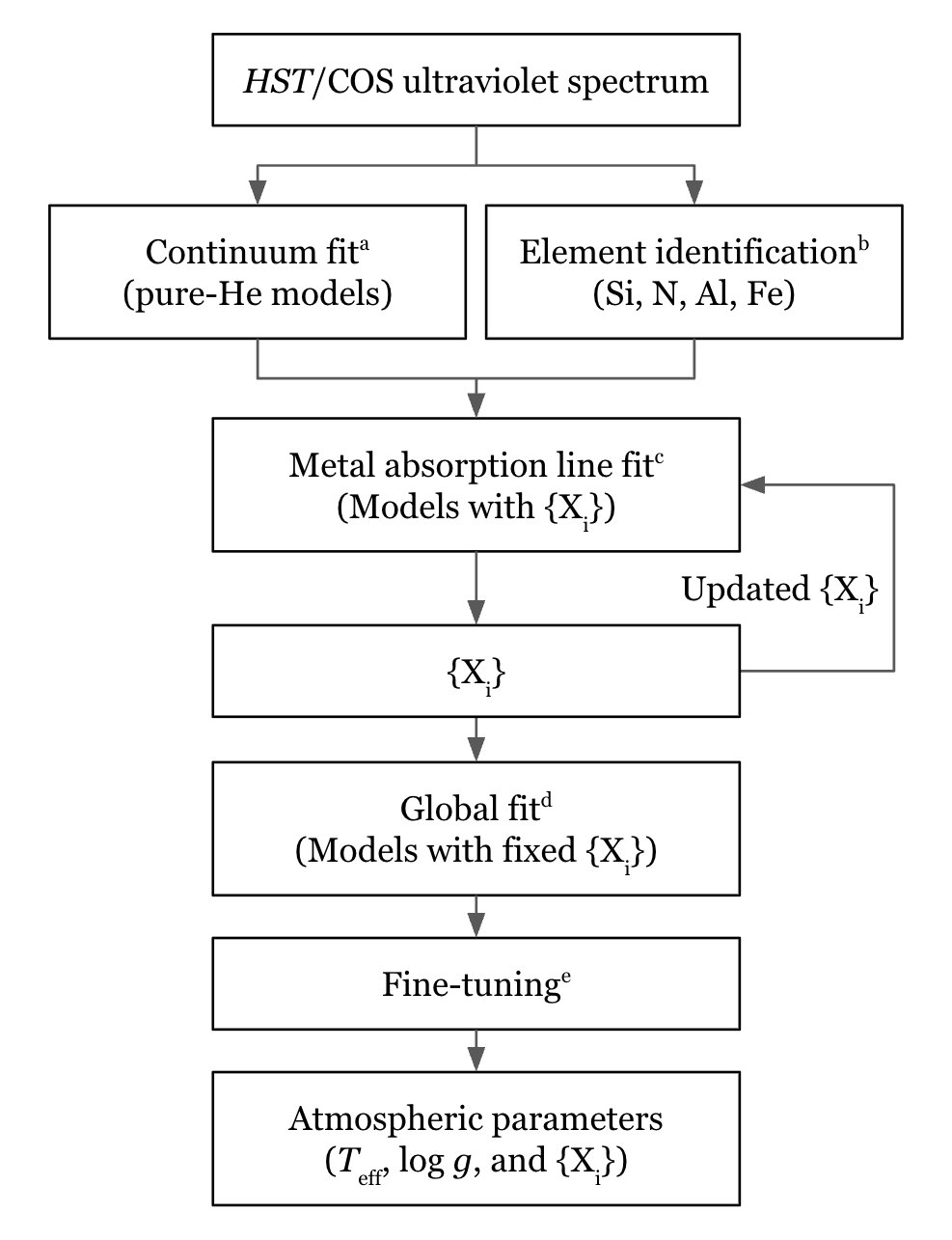}
                \caption{Flowchart of the spectral fitting procedure. Superscripts in the step titles correspond to the steps described in Sect.~\ref{sub.spectral_fit}.}
                \label{fig.flowchart}
            \end{figure}

            We employed the hydrogen-deficient stellar atmosphere models described by \cite{Koester_2010}. This framework enables us to vary the abundance of individual elements, denoted $X_i$, which we define as the base-10 logarithmic number ratio relative to helium, that is, $X_i = \log (N_i / N_{\mathrm{He}})$, where $N$ denotes the particle number. The chemical profile, denoted $\{X_i\}$, therefore represents the atmospheric abundances of all identified elements. Based on the absence of hydrogen lines in the optical spectra \citep{van_Roestel_2022} and the lack of an identifiable Ly$\alpha$ broad absorption feature in the UV (Fig.~\ref{fig.ztfj2252_spec}), we adopt a conservative lower limit for the hydrogen abundance of $X_\mathrm{H} = -9$. By generating grids with different $T_{\mathrm{eff}}$, $\log g$, and $\{X_i\}$, we constrain the overall atmospheric properties. The model spectra provide the Eddington flux at the stellar surface, \(H_\lambda\). To compare the model with the observed flux, \(F_{\mathrm{obs},\lambda}\), we applied the scaling:
            \begin{equation}
            \label{eq:flux_obs}
            F_{\mathrm{obs},\lambda} = \left( \frac{4\pi R^2}{d^2} \right) H_\lambda(T_{\mathrm{eff}}, \log g, \{X_i\})~,
            \end{equation}
            where \(R\) is the WD radius and \(d\) is the distance to the system. The radius was computed using the mass--radius relation of the Montr\'eal white dwarf cooling models\footnote{\url{https://www.astro.umontreal.ca/~bergeron/CoolingModels/}}, based on the evolutionary models of \citet{Bédard_2020}. A 16\% bright-spot flux contribution was estimated from the light-curve fit (Sect.~\ref{sub.lc_analysis}). As no model for the bright-spot emission is available, this contribution was included in the spectral modelling as a secondary flat continuum component in $F_\lambda$. We also included the radial velocity $v_\mathrm{rad}$ as a free parameter to account for wavelength shifts from the WD reflex motion, averaged over the \textit{HST} exposure time. We describe below the procedure used to measure the atmospheric parameters of the accreting WD in ZTF\,J2252$-$05 by fitting its averaged UV spectrum. All spectral fits were carried out using the Markov Chain Monte Carlo (MCMC) method, implemented via the \texttt{emcee} Python package \citep{Foreman-Macke_2013, Hogg_2018}. An overview of the methodology is also presented in the flowchart shown in Fig.~\ref{fig.flowchart}.

            \textit{(a) Continuum fit.}  
            We performed the initial fit on the spectrum's continuum using a grid of pure-helium (DB-type) WD atmosphere models, with  $T_{\mathrm{eff}}$ from 10\,000 to 30\,000\,K in steps of 500\,K, and $\log g$ from 7.5 to 9.5\,dex in steps of 0.2\,dex. Extinction was treated following the approach described in Sect.~\ref{sub.ISM}. This step provided a zeroth-degree estimate of $T_{\mathrm{eff}}$, which was then used as the starting point for the forthcoming spectral fits. 
        
            \textit{(b) Element identification.}  
            As shown in Fig.~\ref{fig.ztfj2252_spec}, the elements with absorption lines suitable for reliable fitting are Si, N, Al, and Fe. The absence of a prominent carbon feature suggests a low abundance. An attempt to constrain its upper limit is presented in Sect.~\ref{sub.carbon_hydrogen_constrain} and is not included in the fitting process.
        
            \begin{table}
                \centering
                \caption{Metal absorption lines used in the spectral fitting.}
                \label{tab:fitted_lines}
                \begin{tabular}{ll}
                    \toprule
                    Element & Fitted lines (\AA) \\
                    \midrule
                    Si  & 1262$^{*}$, 1530$^{*}$ \\
                    N   & 1167, 1493$^{*}$ \\
                    Al  & 1723$^{*}$, 1764 \\
                    Fe  & 1576$^{\dagger}$, 1694$^{\dagger}$ \\
                    \bottomrule
                \end{tabular}
                \tablefoot{An asterisk ($^{*}$) marks doublets. A dagger ($^{\dagger}$) indicates blends of nearby transitions.}
            \end{table}

            \textit{(c) Individual abundance estimation.}
            Due to computational limitations, we did not vary $T_{\mathrm{eff}}$, $\log g$, and abundances simultaneously. At the early stages, $\log g$ was fixed to the value derived from the continuum fit, and model grids were used to explore combinations of $T_{\mathrm{eff}}$ and abundances. For example, Si lines were fitted using grids with varying $T_{\mathrm{eff}}$ and $X_\mathrm{Si}$ while holding $\log g$ constant. In each iteration, the updated $\{X_i\}$ from the previous step served as the background, gradually narrowing the parameter space. To minimise blending effects, we restricted the fits for each element to wavelength regions with relatively isolated absorption lines. The fitted lines are listed in Table~\ref{tab:fitted_lines}. This iterative procedure forms the basis for determining the complete chemical profile of ZTF\,J2252$-$05.
        
            \textit{(d) Global fit.}  
            Using the chemical profile estimated from the previous step, we constrained $T_{\mathrm{eff}}$ and $\log g$ by fitting the full spectrum with grids of models at fixed metal abundances. We masked regions affected by geocoronal Ly$\alpha$ contamination and by potential accretion disc emission features (\ion{N}{V} and \ion{Si}{IV}). The $\lambda1175$\,\AA, $\lambda1549$\,\AA, and $\lambda1640$\,\AA\ lines, which likely originated from the disc or from veiling gas, were not included in the fit.\footnote{In eclipsing accreting binaries, the line of sight can intercept veiling material extending above the disc, giving rise to additional absorption features. An example is the eclipsing system OY~Car, where \citet{Horne_1994} reported a blend of \ion{Fe}{II} features. Samples of eclipsing cataclysmic variables exhibiting similar behaviour were also discussed by \citet{Pala_2017,Pala_2022}.}

            \textit{(e) Fine-tuning.} 
            Due to line blending, neighbouring absorption lines can influence one another, causing the atmospheric parameters estimated from the global fit (i.e. $T_{\mathrm{eff}}$ and $\log g$) to differ slightly from those obtained through initial individual abundance estimations. To resolve this inconsistency, we repeat the individual abundance fitting using the globally determined $T_{\mathrm{eff}}$ and $\log g$. The resulting updated chemical profile is then incorporated into a new global fit. This process is iterated until all parameters converge, completing the fitting procedure.

        \subsubsection{Uncertainty estimation} \label{sub.uncertainty}

            During the spectral fitting process, the distance and extinction were fixed, and their associated systematic uncertainties were not initially considered. Interstellar extinction affects the slope of the spectral continuum, thereby influencing the measurement of $T_{\mathrm{eff}}$. As shown in Eq.~\ref{eq:flux_obs}, the distance influences the overall flux level quadratically, which in turn affects the measurement of $\log g$. To estimate the systematic uncertainties on the atmospheric parameters, we follow a Monte Carlo approach to independently assess the influence of varying $T_{\mathrm{eff}}$ and $\log g$. For extinction, we draw random values from a Gaussian distribution defined by the measured $A_V$ and its uncertainty to perform spectral fitting using $\chi^2$ minimisation over 5000 iterations. The resulting posterior distributions reflect the systematic uncertainty due to extinction. A similar procedure is repeated for the uncertainty in the distance. These systematic uncertainties are then combined in quadrature with the statistical uncertainties obtained from the MCMC analysis. The final atmospheric parameters and uncertainties are listed in Table~\ref{tab:system_parameters}.

    \subsection{Carbon and hydrogen abundance constrain}
    \label{sub.carbon_hydrogen_constrain}
    
        As discussed in Sect.~\ref{sub.ISM}, the \ion{C}{II}\,$1335$\,\AA\ line is affected by interstellar absorption. No other carbon feature in the observed spectrum is strong enough for a reliable abundance fit, suggesting a carbon-poor atmosphere. However, given the critical role of carbon in nuclear burning and as a tracer of evolutionary channels \citep{Nelemans_2010, Toloza_2023}, we attempted to place an upper limit on its abundance using the relatively uncontaminated spectral region near the theoretical \ion{C}{I} line at $1280$\,\AA. We simulated a grid of models based on the best-fit spectrum with varying carbon abundances. These were then compared with the observed spectrum of ZTF\,J2252$-$05 to determine an upper limit on the carbon abundance. Hydrogen is also a key diagnostic for identifying systems that evolved through the CV channel \citep{Podsiadlowski_2003, Belloni_2023}. To test for its presence, we carried out a similar exercise for hydrogen by increasing its abundance stepwise from a conservative lower limit.
        


\section{Results} \label{Result}
    
\begin{figure*}
            \centering
            \begin{subfigure}[b]{0.48\textwidth}
                \includegraphics[width=\textwidth]{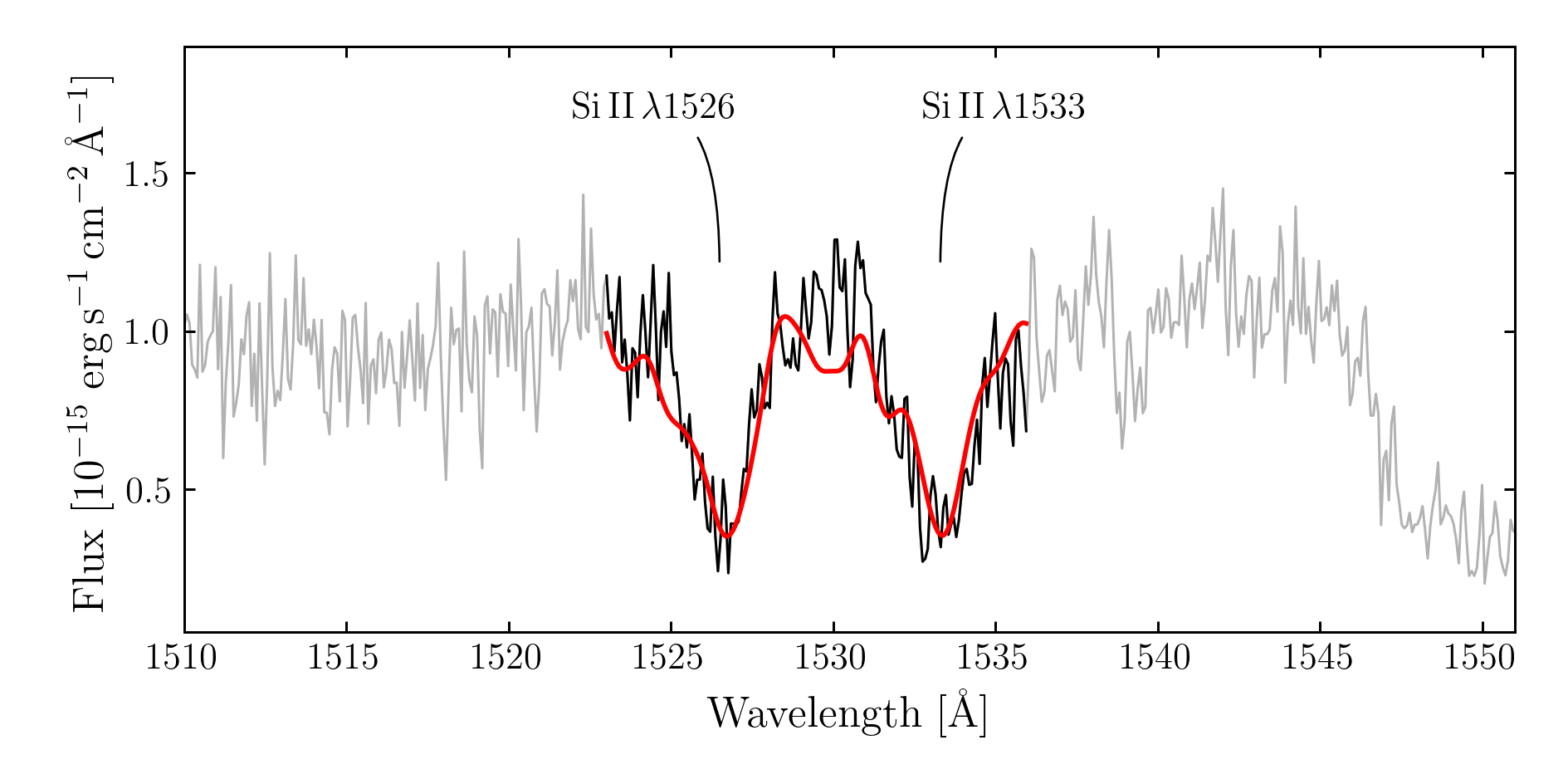}
            \end{subfigure}
            \hfill
            \begin{subfigure}[b]{0.48\textwidth}
                \includegraphics[width=\textwidth]{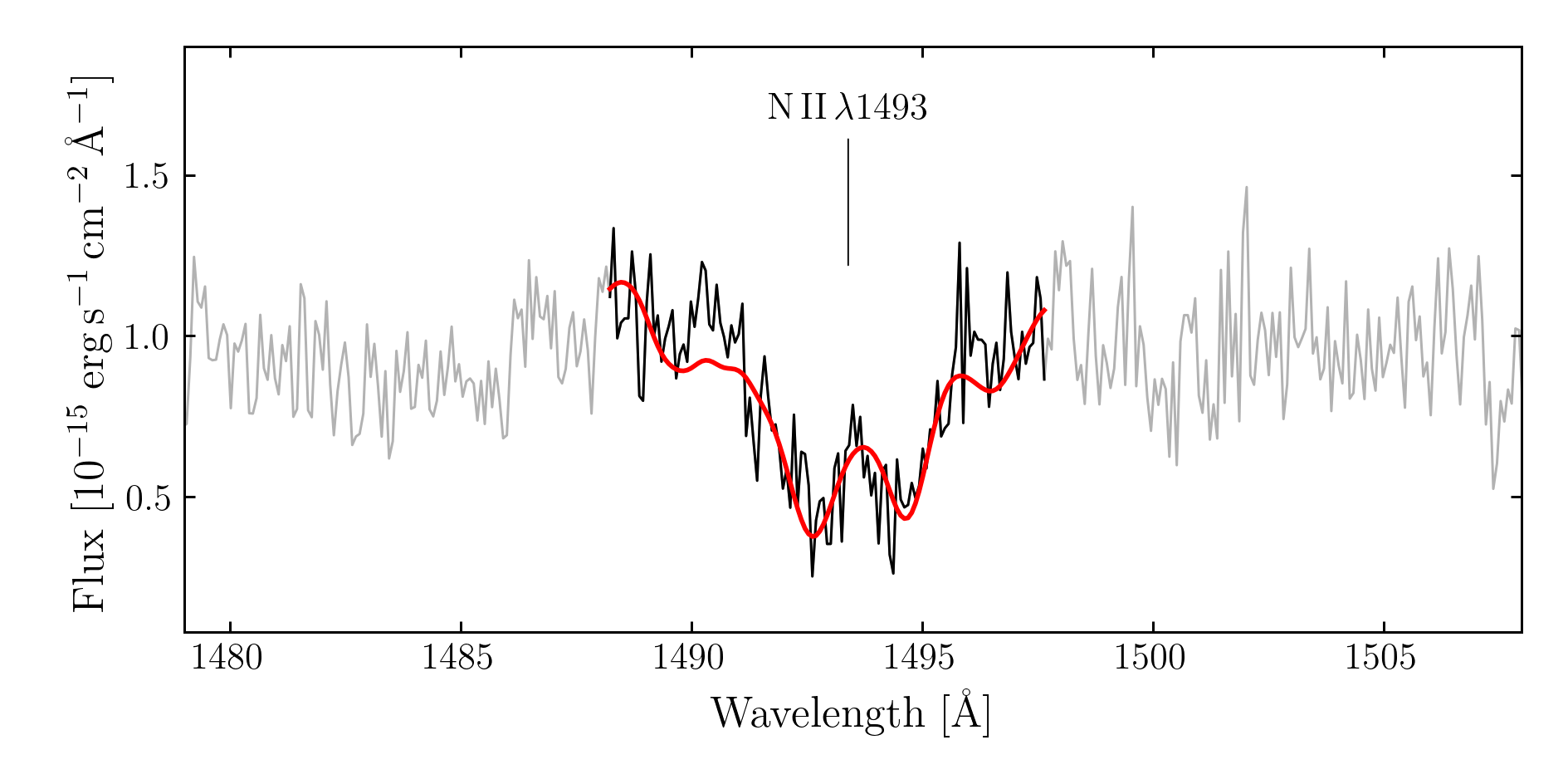}
            \end{subfigure}
        
            \vspace{0.2em}
        
            \begin{subfigure}[b]{0.48\textwidth}
                \includegraphics[width=\textwidth]{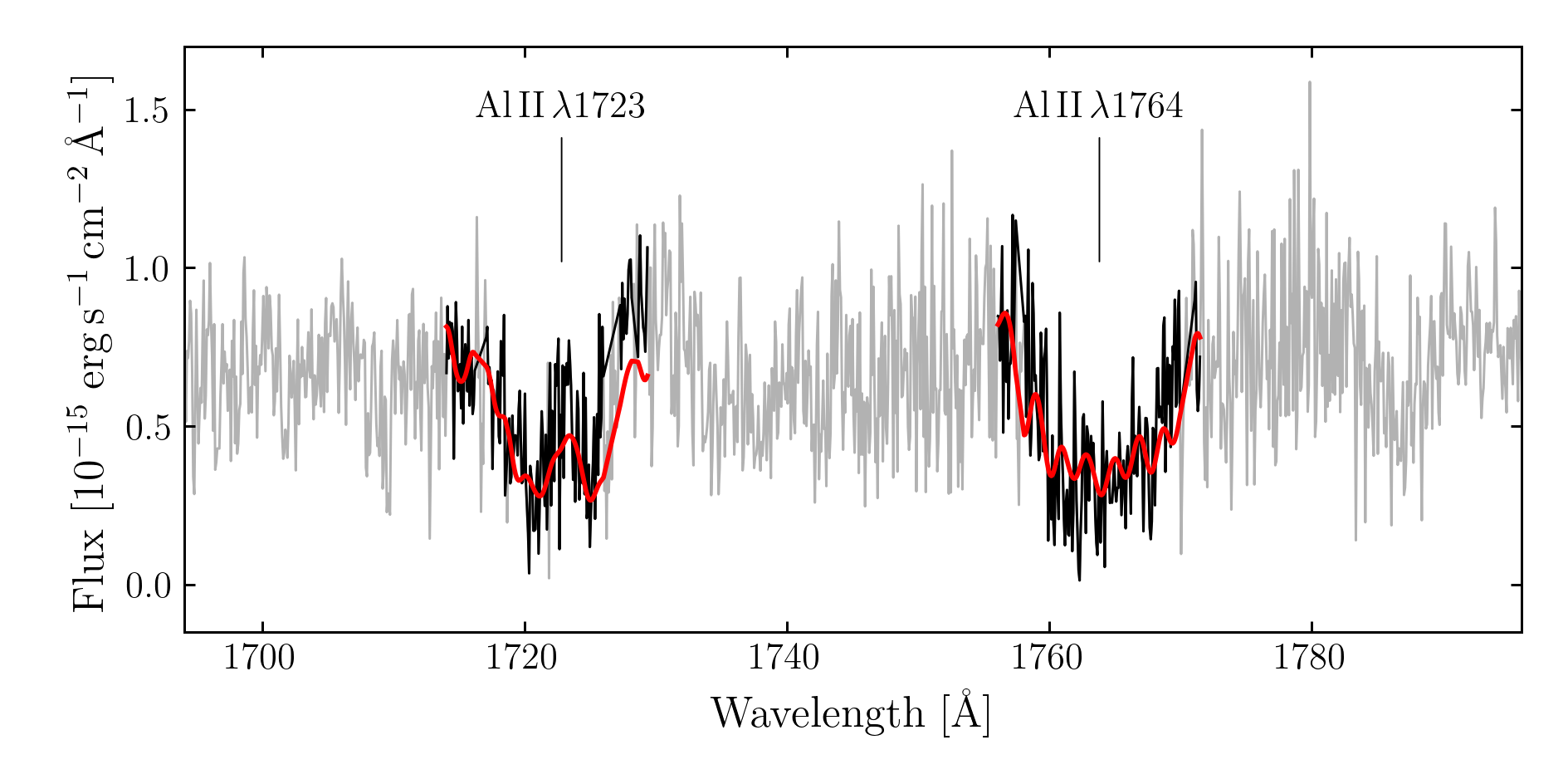}
            \end{subfigure}
            \hfill
            \begin{subfigure}[b]{0.48\textwidth}
                \includegraphics[width=\textwidth]{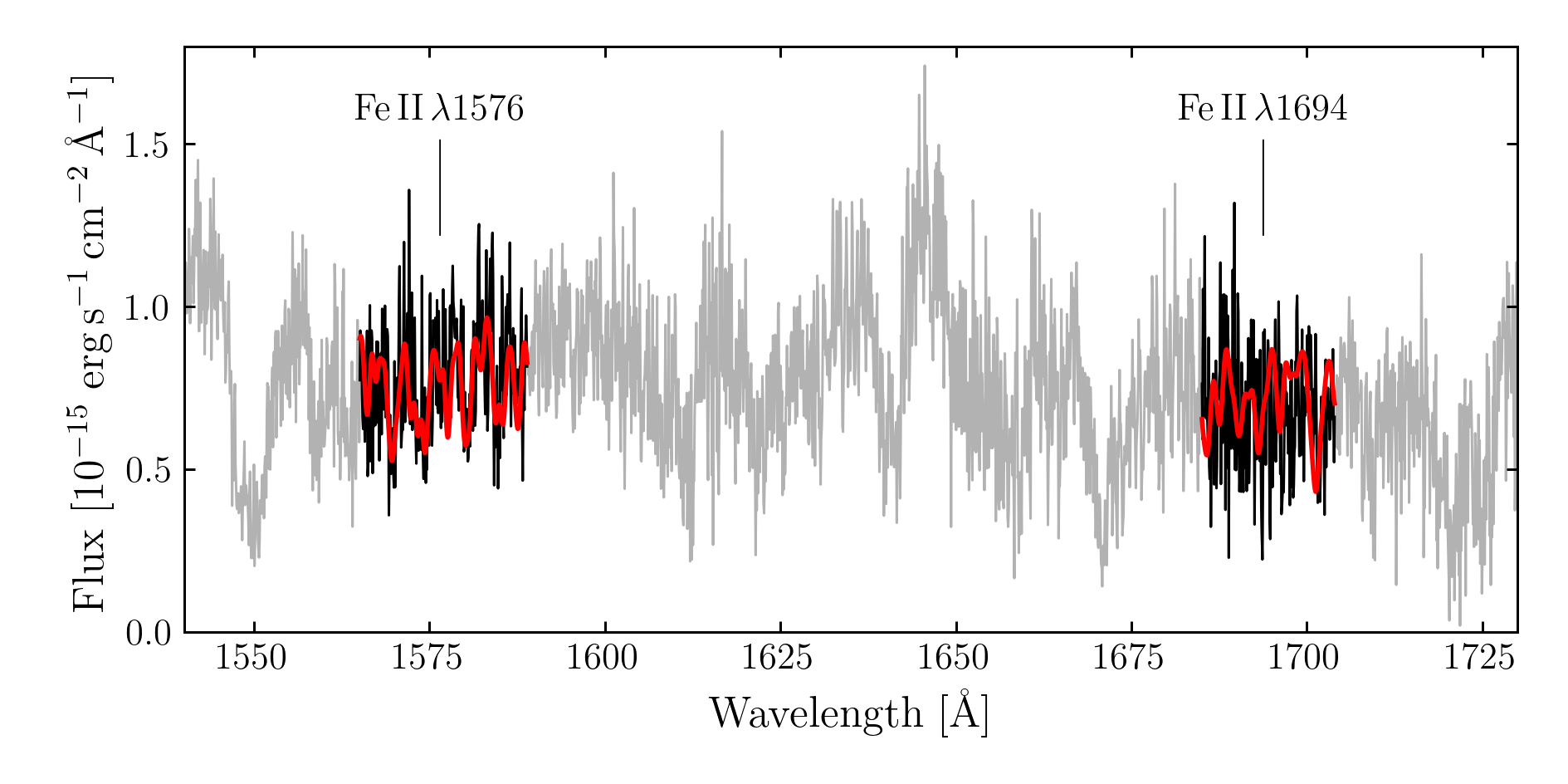}
            \end{subfigure}
            
            \caption{Spectral fits for the individual elements Si, N, Fe, and Al (panels arranged clockwise from top left). This corresponds to step~(c) described in Sect.~\ref{sub.spectral_fit}. The identified absorption lines (black) are compared with the best-fitting models (red), while spectral regions not included in the fit are shown in grey.}
            \label{fig.spectral_fit_individual}
    \end{figure*}

\begin{figure*}
        \centering
        \includegraphics[width=\textwidth]{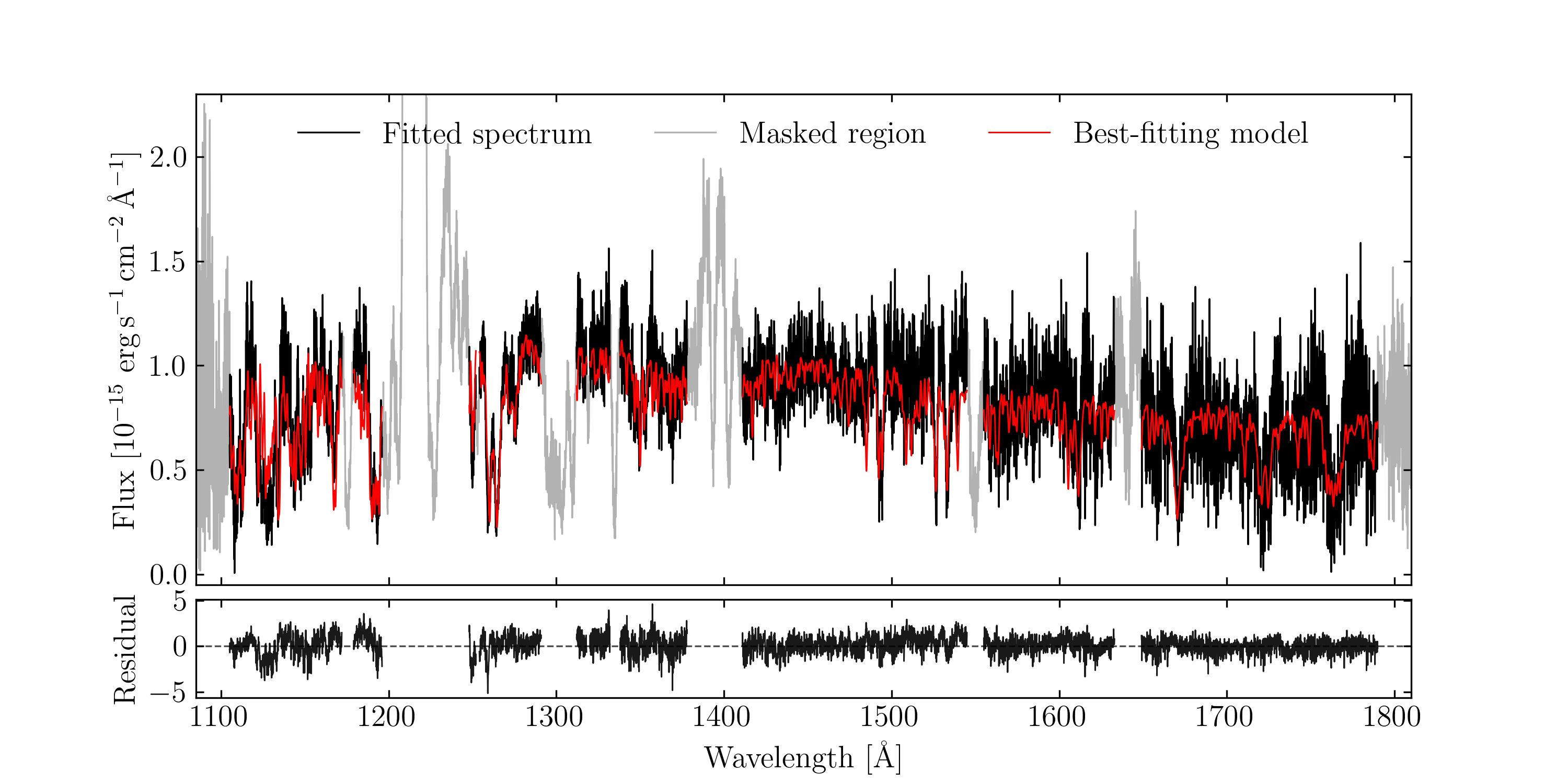}
        \caption{Best-fitting atmospheric model for ZTF\,J2252$-$05, computed with the hydrogen-deficient code of \citet{Koester_2010} using the parameters listed in Table~\ref{tab:system_parameters}. The bright spot contribution is included as an additional constant term in $F_\lambda$, accounting for about $16\,\%$ of the total flux. The model includes Si, N, Al, and Fe. Geocoronal lines, ISM-contaminated regions, potential disc emission, and possible veiling absorption were masked. The residuals are shown below the spectrum.}
        \label{fig.spectral_fit_best}
    \end{figure*}

    Figure~\ref{fig.spectral_fit_individual} shows the individual atmospheric models fitted for the identified elements (Si, N, Al, and Fe). These abundances are included in the global fit over the studied wavelength range, and the resulting best-fit stellar atmosphere model is shown in Fig.~\ref{fig.spectral_fit_best}. The parameters derived from the spectral fits are listed in Table~\ref{tab:system_parameters}. 

\begin{figure*}
        \centering
        \includegraphics[width=0.49\textwidth]{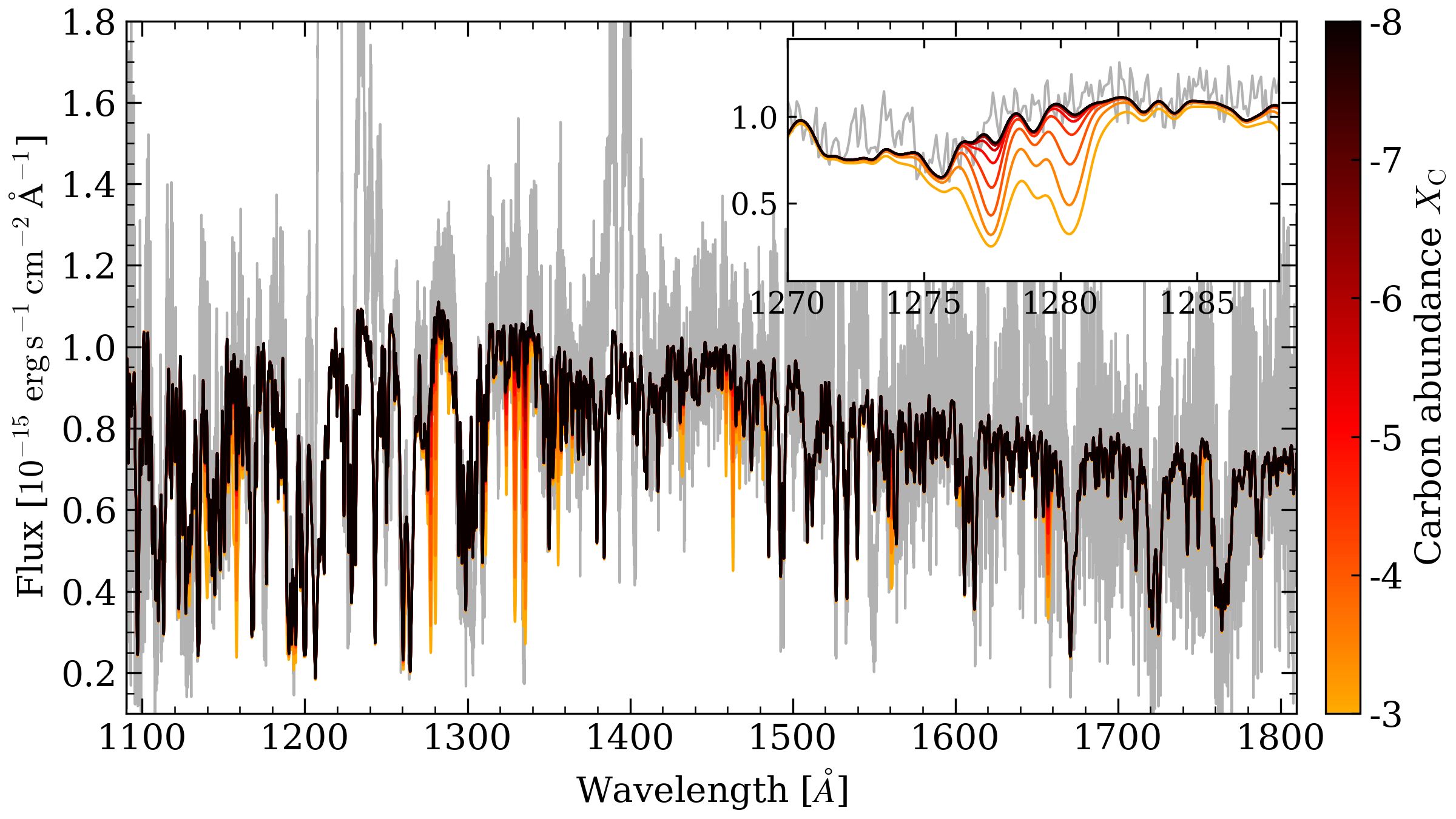}
        \includegraphics[width=0.49\textwidth]{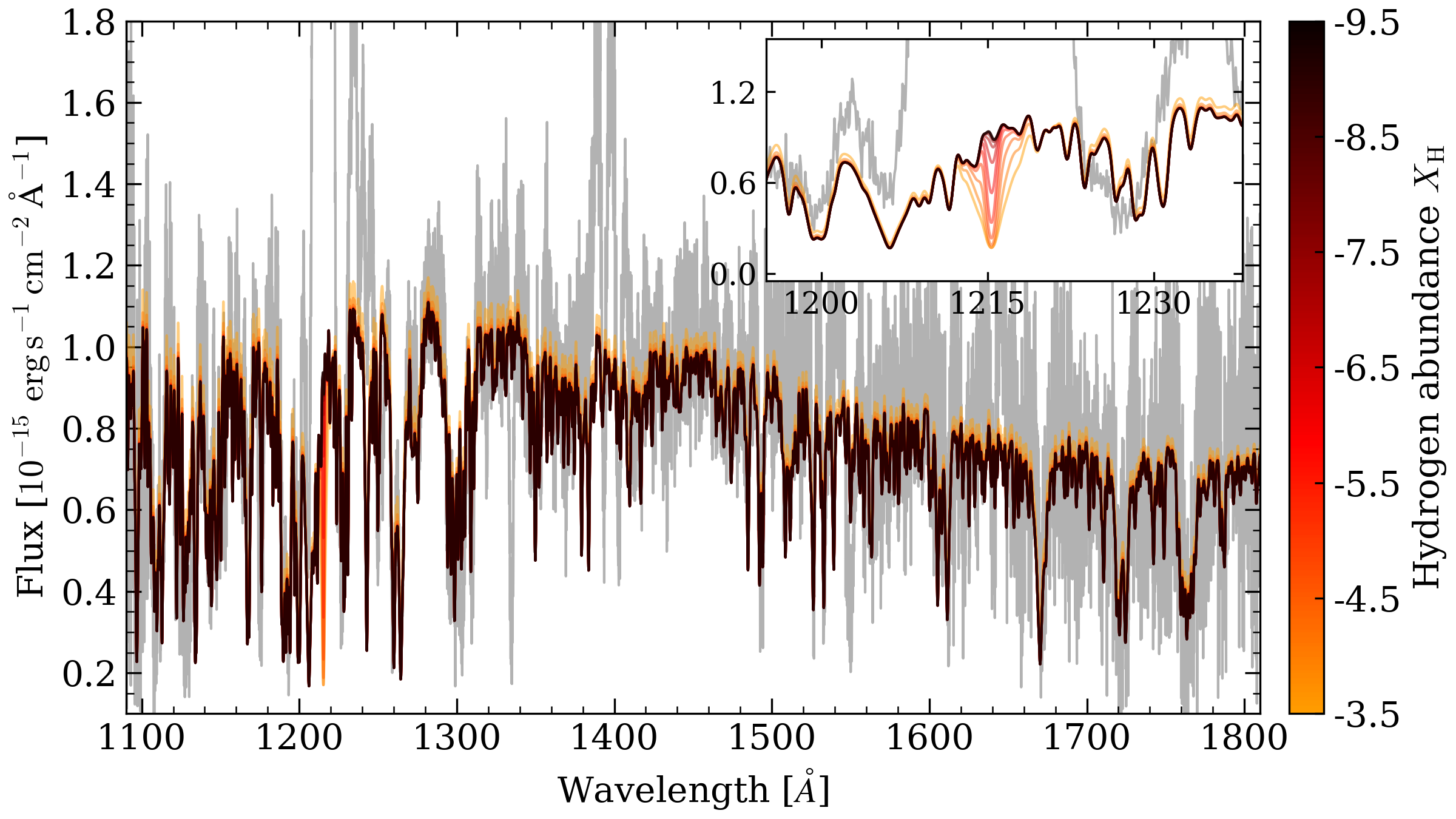}
        \caption{Synthetic spectra with varying carbon (left) and hydrogen (right) abundances overplotted on the observed spectrum (grey). Inset panels highlight the diagnostic regions. \textit{Left:} Carbon abundances ranging from $-8.0$ to $-3.0$ are examined; the absence of C\,\textsc{i} lines near 1280\,\AA\ indicates $X_\mathrm{C} < -5.0$. \textit{Right:} Hydrogen abundances from $-9.5$ to $-3.0$ are tested; Ly$\alpha$ is outshone by geocoronal emission, and no clear hydrogen diagnostic lines are present, leaving the hydrogen abundance unconstrained.}
        \label{fig.C_H_estimation}
    \end{figure*}

    The synthetic spectra with varying carbon abundances, based on the best-fit model, are shown in Fig.~\ref{fig.C_H_estimation} (left panel). The colour bar indicates the carbon abundance, ranging from $X_\mathrm{C} = -3.0$ to $X_\mathrm{C} = -8.0$ in steps of 0.5. By comparing the models with the expected \ion{C}{I} absorption feature near 1280\,\AA, we derived an upper limit of $X_\mathrm{C} < -5.00$.

    We fixed a conservative hydrogen abundance of $X_\mathrm{H} = -9$ in the global spectral fit, as no hydrogen features were detected in the optical spectroscopic analysis of ZTF\,J2252$-$05 \citep{van_Roestel_2022}. In the UV, the Ly$\alpha$ line at 1215\,\AA\ is the most prominent feature but is completely outshone by geocoronal emission. Figure~\ref{fig.C_H_estimation} (right panel) shows models with hydrogen abundances ranging from $X_\mathrm{H} = -9.5$ to $X_\mathrm{H} = -3.0$ in steps of 0.5. Variations in $X_\mathrm{H}$ primarily affect the continuum, leading to degeneracies with surface gravity. No isolated hydrogen absorption line was available for a reliable estimate of $X_\mathrm{H}$, and the hydrogen abundance therefore remains unconstrained.

    For the primary WD, the spectral fit yields an effective temperature of $T_{\mathrm{eff}} = 23\,300 \pm 600$\,K and a surface gravity of $\log g = 8.4 \pm 0.3$. Using the mass--radius relation from \citet{Bédard_2020}, we derive a mass of $M_{\mathrm{WD}} = 0.86 \pm 0.16\,M_\odot$ and a radius of $R_{\mathrm{WD}} = 0.0095 \pm 0.0018\,R_\odot$. For the donor star, we adopted a mass ratio of $q = 0.034 \pm 0.006$ from the photometric analysis of \citet{van_Roestel_2022}, which implies a donor radius of $R_{\mathrm{donor}} = 0.049 \pm 0.004\,R_\odot$ under the assumption of a Roche-lobe-filling configuration. This yields a donor mass of $M_{\mathrm{donor}} = 0.029 \pm 0.008\,M_\odot$, from which we calculate an orbital separation of $a = 0.36 \pm 0.02\,R_\odot$.

    \begin{table}
        \caption{System parameters of ZTF\,J2252$-$05.}
        \label{tab:system_parameters}
        \centering
        \normalsize
        \begin{tabular}{l r}
            \toprule
            Parameter & Value \\
            \midrule
            $d^{*}$ [pc]                      & $536^{+82}_{-93}$ \\
            $A_V$ [mag]                       & $0.149 \pm 0.007$ \\
            $P_{\mathrm{orb}}^{*}$ [min]      & 37.3941123(8) \\
            $q^{*}$                           & $0.034 \pm 0.006$ \\
            $i^{*}$ [deg]                         & $87 \pm 1$ \\
            $a$ [$R_\odot$]                  & $0.36 \pm 0.02$ \\
            \addlinespace
            $T_{\mathrm{eff}}$ (K)            & $23300\pm600$  \\
            $\log g$                          & $8.4\pm0.3$  \\
            $X_\mathrm{Si}$                   & $-3.92\pm0.13$ \\
            $X_\mathrm{N}$                    & $-2.58\pm0.11$ \\
            $X_\mathrm{Al}$                   & $-3.73\pm0.12$ \\
            $X_\mathrm{Fe}$                   & $-3.26\pm0.10$ \\
            $X_\mathrm{C}$                    & $<-5.00$       \\
            $v_{\mathrm{rad}}$ (km\,s$^{-1}$) & $54\pm3$ \\
            $M_{\mathrm{WD}}$ [$M_\odot$]     & $0.86 \pm 0.16$ \\
            $R_{\mathrm{WD}}$ [$R_\odot$]     & $0.0095 \pm 0.0018$ \\
            \addlinespace
            $M_{\mathrm{donor}}$ [$M_\odot$]  & $0.029 \pm 0.008$ \\
            $R_{\mathrm{donor}}^{*}$ [$R_\odot$]  & $0.049 \pm 0.004$ \\
            \bottomrule
        \end{tabular}
        \vspace{2mm}
        \tablefoot{Values adopted from \citet{van_Roestel_2022} are marked with an asterisk ($^{*}$). $R_{\mathrm{donor}}$ is set by the mass ratio $q$, assuming a Roche-lobe filling donor.} \\
    \end{table}

\section{Discussion} \label{Discussion}

    \subsection{Formation channel identification} \label{sub.formation_channel}

        \subsubsection{Chemical abundances}
    
            In Sect.~\ref{Introduction}, we outlined the three main evolutionary channels for AM\,CVn systems: WD channel, the semi-degenerate He-star channel, and the CV channel. \citet{Nelemans_2010} proposed a method to infer the likely formation pathway based on the observed chemical abundances in the accreting WD's atmosphere originating from the donor star. Their approach combines stellar evolution models \citep{Eggleton_2002} with magnetic braking prescriptions \citep{Rappaport_1983, Podsiadlowski_2002}, showing that the chemical signature of the accreted material traces the evolutionary state of the donor star.
        
\begin{figure}
                \centering
                \includegraphics[width=\columnwidth]{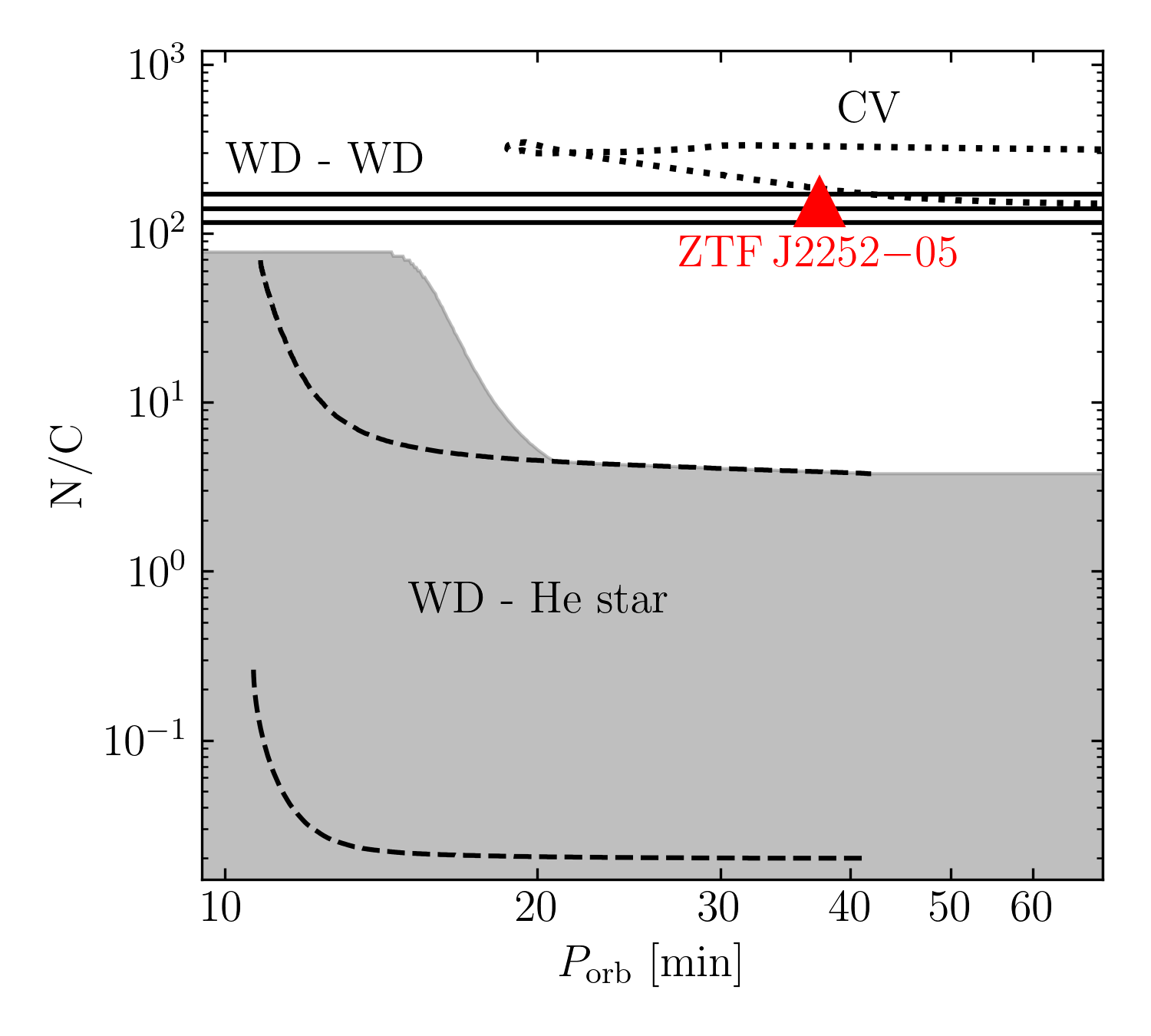}
                \caption{The derived limit N/C $> 153$ (red triangle) is compared with evolutionary models. WD and He-star channel models are taken from \citet{Nelemans_2010}. White-dwarf donor tracks with progenitor masses of 1.0, 1.5, and 2.0\,$M_\odot$ (from top to bottom) are shown as solid lines. The possible He-star donor region is indicated by the shaded area, with two examples for initial orbital periods of 20 and 60\,min (upper and lower dashed lines). The CV-channel model (dashed line) is calculated using system parameters consistent with Table~\ref{tab:system_parameters}.}
                \label{fig.abundance_ratios}
            \end{figure}

            The detection of hydrogen in the spectrum was long regarded as strong evidence for the CV channel (e.g. \citealt{Podsiadlowski_2003, Solheim_2010}), but \citet{Belloni_2023} recently showed that this channel can also produce AM\,CVn systems with undetectable hydrogen abundances ($X_\mathrm{H} \lesssim -6$). Abundances of elements heavier than helium are likely to be similar between the WD and CV channels \citep{Nelemans_2010}. However, distinguishing the He-star channel from other channels is still possible with a detailed assessment of the atmospheric metal abundances. In particular, when material is processed through the CNO cycle, the ratios of carbon, nitrogen, and oxygen evolve in characteristic patterns. Helium burning leads first to the production of carbon, followed by oxygen, effectively enhancing their abundances relative to that of nitrogen. This makes the nitrogen-to-carbon abundance ratio N/C a useful diagnostic: systems formed via the WD or CV channel tend to show a high N/C, while those from the He-star channel show lower values due to ongoing helium burning. Figure~\ref{fig.abundance_ratios} illustrates this diagnostic using model predictions of the N/C evolution for various binary configurations.
        
            From our spectrum, we estimate $\log(\mathrm{N}/\mathrm{He}) = -2.58$ and $\log(\mathrm{C}/\mathrm{He}) < -5.00$ (by particle number). Converting these results to a mass ratio using the atomic masses of nitrogen and carbon ($m_\mathrm{N} = 14.007\,u$ and $m_\mathrm{C} = 12.011\,u$; $1\,u = 1.6605\times10^{-24}\,\mathrm{g}$), we obtain a lower limit of $\mathrm{N}/\mathrm{C} > 153$. This value is compared with theoretical predictions in Fig.~\ref{fig.abundance_ratios}, placing ZTF\,J2252$-$05 well above the range spanned by the He-star models and strongly supporting a WD or CV formation scenario.
        
        \subsubsection{Stellar evolutionary models}
\begin{figure}
                \centering
                \includegraphics[width=\columnwidth]{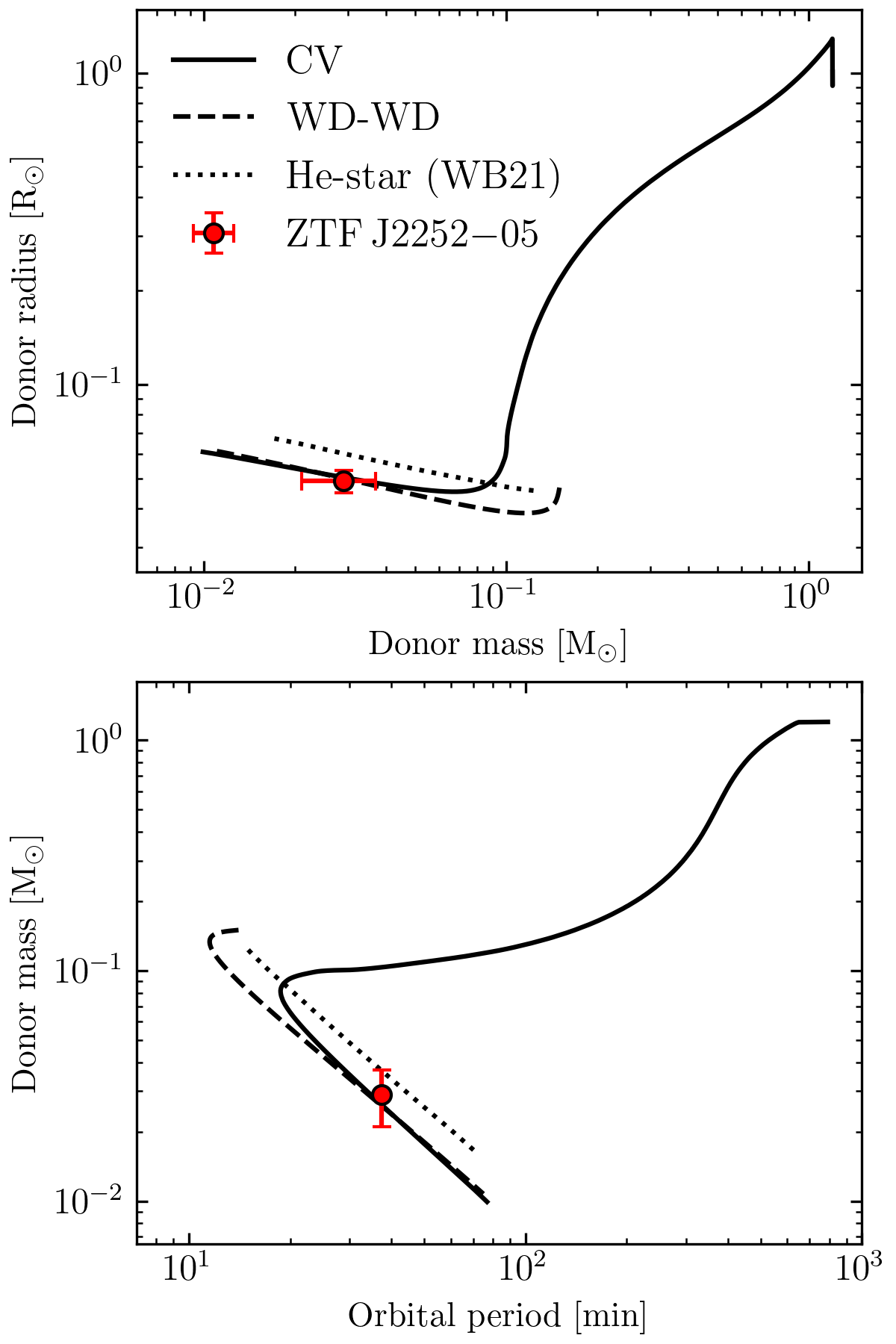}
                \caption{Comparison between the measured donor properties and theoretical evolutionary tracks computed with \texttt{MESA}. The upper panel shows the donor radius $R_\mathrm{donor}$ as a function of donor mass $M_\mathrm{donor}$; the lower panel shows the corresponding relation between $M_\mathrm{donor}$ and orbital period. The CV channel (solid line) and the adiabatic white dwarf channel (dashed line) reproduce the spectroscopically derived parameters, while an example He-star channel track (dotted line) from \citet{Wong_2021}, labeled “WD21”, is included for reference.}
                \label{fig.MESA_R_M}
            \end{figure}

            We compare our estimated parameters with theoretical AM\,CVn evolutionary models computed using the stellar evolution software Modules for Experiments in Stellar Astrophysics \citep[\texttt{MESA};][]{Paxton_2011, Paxton_2013, Paxton_2015, Paxton_2018, Paxton_2019}. In Fig.~\ref{fig.MESA_R_M}, we present the evolution of $R_\mathrm{donor}$ and $M_\mathrm{donor}$ for the three proposed AM\,CVn formation channels. 
            
            The model for the CV channel was computed following \citet{Belloni_2023}. We ran a grid of models varying the initial $M_{\mathrm{donor}}$ and the initial $P_{\mathrm{orb}}$, assuming an initial $M_{\mathrm{WD}}$ of $0.86\,M_\odot$. Depending on the mass transfer rate, the WD mass can increase, and we adopted the mass growth criteria from \citet{Wolf_2013}. The value of $M_{\mathrm{donor}}$ is varied between $1.0$ and $1.5\,M_\odot$ in steps of $0.1\,M_\odot$, and the value of $P_{\mathrm{orb}}$ between $0.2$ and $2.0$ days in steps of $0.2$ days. The grid is then refined around the region where models reproduce the properties derived from the observational data.
            
\begin{figure}
                \centering
                \includegraphics[width=\columnwidth]{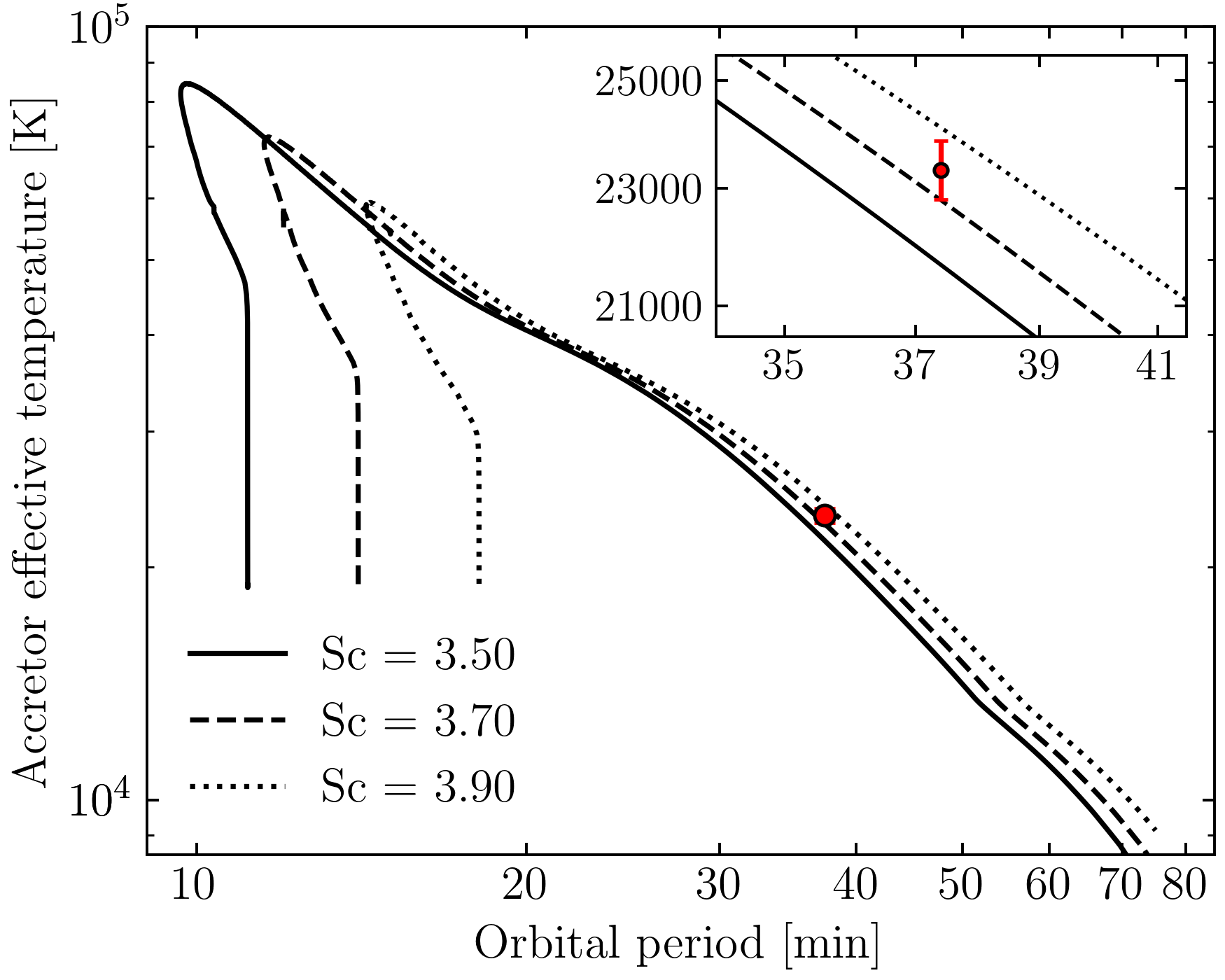}
                \caption{Comparison between the estimated WD effective temperature $T_\mathrm{eff}$ and theoretical WD channel models with varying central entropy $S_\mathrm{C}$.}
                \label{fig.MESA_teff}
            \end{figure}

            The adiabatic WD channel models were produced following \citet{Wong_2021}. For each value of central specific entropies $S_\mathrm{c}$, we rapidly reduce the mass of a 0.15\,$M_\odot$ He WD (taken from \citealt{Wong_2023}) via a high mass-loss rate to ensure adiabatic evolution, and measure its radius at various points. In this channel, the evolution is governed directly by the donor’s $S_\mathrm{c}$, which sets the mass–radius relation \citep{Deloye_2007}, whereas in the CV channel the donor entropy is fixed by the helium core mass at the onset of mass transfer \citep{Belloni_2023}. From the adiabatic mass–radius relation, we numerically integrate the orbital evolution equations assuming fully conservative mass transfer and angular momentum loss driven solely by gravitational wave radiation (e.g. \citealt{Bauer_2021}). The resulting mass transfer rate is then applied to a CO WD accretor with an initial mass between 0.65 and 0.85\,$M_\odot$ to model its accretion-induced reheating and subsequent cooling. The evolution of $T_\mathrm{eff}$ with $P_\mathrm{orb}$ is shown in Fig.~\ref{fig.MESA_teff}.
        
            The CV and WD–WD evolutionary models reproduce our spectroscopically derived parameters within their uncertainties, indicating consistency between theoretical predictions and observations. Although it is unlikely that this system formed through the He-star channel, for completeness, we present the corresponding donor parameters using models from \citet{Wong_2021}. This model assumes an adiabatic evolution after the mass of the He-star donor decreases to 0.2 $M_\odot$. As the donor masses of all three tracks overlap considerably within the typical AM\,CVn $P_\mathrm{orb}$ range of 5--70\,min, identifying the formation channel based on this criterion alone remains challenging. However, this method becomes more promising for systems with shorter $P_\mathrm{orb}$, such as ES\,Cet, KIC\,4547333, and HP\,Lib (see Table~\ref{tab:HST_program}). Therefore, chemical analysis remains a crucial tool for constraining the formation pathway.
    
    \subsection{Comparison with optical results}
    \label{sub.compare_with_optical}

        The mass and radius derived from our spectral fit of the primary WD are consistent with the values reported by \citet{van_Roestel_2022}, who obtained $M_{\mathrm{WD}} = 0.76 \pm 0.05\,M_\odot$ and $R_{\mathrm{WD}} = 0.010 \pm 0.001\,R_\odot$. In contrast, the effective temperature inferred from our fit is significantly higher than their reported value of \(T_{\mathrm{eff}} = 15\,200 \pm 900\) K. A similar exercise by \citet{Macrie_2024} yielded \(T_{\mathrm{eff}} = 15\,560 \pm 460\) K. This discrepancy likely reflects differences in the modelling. The earlier studies employed a simplified blackbody fit and adopted a lower extinction value of \(E(B-V) = 0.01\) from \citet{Green_2019}. This interstellar dust map does not account for sources within 400\,pc and likely underestimates the total extinction. Whereas our ISM model, based on the more recent 3D dust map from \citet{Lallement_2022}, yields $E(B-V) = 0.048$ and enables a more accurate spectral modelling.

\section{Summary and conclusion} \label{Conclusion}

    We analysed \textit{HST}/COS time-tagged UV spectroscopy of the AM\,CVn system ZTF\,J2252$-$05. This system was selected as a benchmark target to demonstrate the feasibility of reconstructing the evolutionary history of AM\,CVns through detailed atmospheric spectral fit. By combining updated astrometry from Gaia DR3 \citep{Brown_2021}, extinction corrections from 3D dust maps \citep{Lallement_2022}, and published photometry and light-curve modelling \citep{van_Roestel_2022}, we constrained the parameters of the accreting white dwarf using hydrogen-deficient atmospheric models with realistic chemical compositions, applied in this context for the first time.  

    For the accretor, we determined an effective temperature of $T_{\mathrm{eff}} = 23\,300 \pm 600~\mathrm{K}$, a surface gravity of $\log g = 8.4 \pm 0.3$, a mass of $M_{\mathrm{WD}} = 0.86 \pm 0.16~M_\odot$, and a radius of $R_{\mathrm{WD}} = 0.0095 \pm 0.0018~R_\odot$. These results highlight the importance of UV spectroscopy for characterising ultracompact binaries. The derived mass is consistent with \citet{van_Roestel_2022} and supports the view that accretors in AM\,CVn binaries are generally more massive than typical single white dwarfs ($\approx0.6\,M_\odot$), instead resembling the masses of accretors in CVs \citep{Zorotovic_2011, Pala_2022}. The measured $T_\mathrm{eff}$ is higher than reported by \citet{van_Roestel_2022}, partly due to underestimated extinction, and further indicates that spectral energy distribution fits with simplified blackbody models may misjudge the accretor temperature, highlighting the need for detailed atmospheric modelling.
    
    We identified absorption features of the accreting WD from Si, N, Al, and Fe, and measured their atmospheric abundances. We also placed an upper limit on the abundance of C, which plays a key role in thermonuclear processes and serves as a tracer of the formation channel. From the measured abundances, we estimated a nitrogen-to-carbon mass ratio of N/C $> 153$. Among the three proposed AM\,CVn formation channels, our results strongly disfavour the He-star channel, while the WD and CV channels remain viable. The presence of hydrogen below the detection limit cannot be excluded, leaving these two channels indistinguishable.
    
    This work develops and validates a robust pipeline for spectral modelling of mass-transferring, hydrogen-deficient ultra-compact binaries. Our results show that atmospheric abundance analysis provides meaningful constraints on the evolutionary origins of individual systems. With this methodology now tested and validated on ZTF\,J2252$-$05, we are well positioned to apply the same approach to a larger sample of AM\,CVn systems. This paves the way toward constructing the first statistically significant spectroscopic population with well-characterised formation channels, placing AM\,CVn stars within the broader landscape of accreting compact binaries.


\begin{acknowledgements}
We thank Lars Bildsten for valuable insights and discussions. We acknowledge with thanks the variable star observations from the AAVSO International Database contributed by observers worldwide and used in this research. We thank the members of the Spanish Observers of Supernovae (ObSN) group for their valuable photometric contributions. This research was supported by Deutsche Forschungsgemeinschaft  (DFG, German Research Foundation) under Germany’s Excellence Strategy - EXC 2121 "Quantum Universe" – 390833306. Co-funded by the European Union (ERC, CompactBINARIES, 101078773). Views and opinions expressed are however those of the author(s) only and do not necessarily reflect those of the European Union or the European Research Council. Neither the European Union nor the granting authority can be held responsible for them. DB acknowledges support from the São Paulo Research Foundation (FAPESP), Brazil, Process Numbers {\#2024/03736-2} and {\#2025/00817-4}. MRS is supported by Fondecyt (grant 1221059). MJG acknowledges support from the European Research Council through ERC Advanced Grant No. 101054731, from the National Aeronautics and Space Administration under grants 80NSSC24K0436, 80NSSC22K0479, and 80NSSC24K0380, and from the National Science Foundation under grant AST-2205736. PJG is supported by NRF SARChI grant 111692. PR-G acknowledges support by the Agencia Estatal de Investigación del Ministerio de Ciencia e Innovación (MCIN/AEI) and the European Regional Development Fund (ERDF) under grant PID2021--124879NB--I00. DS is supported by the UK Science and Technology Facilities Council (STFC, grant numbers ST/T007184/1, ST/T003103/1, and ST/T000406/1). OT acknowledges Proyectos Internos USM 2025, PI-LII-2025-03. GT was supported by grants IN109723 from the Programa de Apoyo a Proyectos de Investigación e Innovación Tecnológica (PAPIIT). This project has received funding from the European Research Council (ERC) under the European Union’s Horizon 2020 research and innovation programme (Grant agreement No. 101020057).
\end{acknowledgements}

\bibliographystyle{aa}
\bibliography{references}

\end{document}